\documentclass[useAMS,usenatbib]{mn2e}
\usepackage{graphicx}
\usepackage{multirow}
\usepackage{color}
\usepackage{array}
\usepackage{amsmath}
\usepackage{amssymb}
\usepackage{mathtools}
\usepackage{ifthen}
\usepackage{hyperref}
\usepackage{breakurl}
\usepackage{tabularx}
\title{Population of post-nova supersoft X-ray sources}
\author[Monika D. Soraisam et al.]{Monika D. Soraisam$^{1}$\thanks{E-mail:monikas@mpa-garching.mpg.de}, Marat Gilfanov$^{1,2}$, William M. Wolf $^{4}$, Lars Bildsten$^{4,5}$\\ 
$^{1}$Max-Planck-Institut f\"{u}r Astrophysik, Karl-Schwarzschild-Str. 1, 85748 Garching, Germany\\
$^{2}$Space Research Institute, Russian Academy of Sciences, Profsoyuznaya 84/32, 117997 Moscow, Russia\\
$^{4}$Department of Physics, University of California, Santa Barbara, CA 93106\\
$^{5}$Kavli Institute for Theoretical Physics, Santa Barbara, CA 93106\\
}

\begin{document}

\date{Received/Accepted}

\pagerange{\pageref{firstpage}--\pageref{lastpage}} \pubyear{2015}

\maketitle

\label{firstpage}

\begin{abstract}
Novae undergo a supersoft X-ray phase of varying duration after the optical outburst. Such transient post-nova supersoft X-ray sources (SSSs) are the majority of the observed SSSs in M31. In this paper, we use the post-nova evolutionary models of Wolf et al. to compute the expected population of post-nova SSSs in M31.  We predict that depending on the assumptions about the WD mass distribution in novae, at any instant there are about 250 -- 600 post-nova SSSs in M31 with (unabsorbed)  0.2--1.0~keV luminosity $L_{\rm x}\geq 10^{36}$~erg/s. Their combined unabsorbed luminosity is of the order of $\sim 10^{39}$~erg/s.  Their luminosity distribution shows significant steepening around $\log(L_{\rm x})\sim 37.7$ -- 38 and becomes zero at $L_{\rm x}\approx 2\times 10^{38}$~erg/s, the maximum $L_{\rm x}$ achieved in the post-nova evolutionary tracks. Their effective temperature distribution has a roughly power law shape with differential slope of ${\approx 4\mbox{--}6}$ up to the maximum temperature of $T_{\rm eff}\approx 1.5\times 10^{6}$~K.

We compare our predictions with the results of the {\it XMM-Newton} monitoring of the central field of M31  between 2006 and 2009. The predicted number of post-nova SSSs exceed the observed number by a factor of ${\approx 2\mbox{--}5}$, depending on the assumed WD mass distribution in novae. This is good agreement, considering the number and magnitude of uncertainties involved in calculations of the post-nova evolutionary models and their X-ray output.  Furthermore, only a moderate circumstellar absorption, with hydrogen column density of the order of $\sim 10^{21}~{\rm cm}^{-2}$, will remove the discrepancy.

\end{abstract}


\section{Introduction}\label{sec:intro}
Novae are  a sub-class of cataclysmic variables showing prominent outbursts. These  systems, composed of a white dwarf (WD) accreting matter from a non-degenerate companion,  are characterised by relatively low mass accretion rate, insufficient for stable nuclear burning of the accreted material. The nova explosion results from a thermonuclear runaway (TNR) which develops under degenerate conditions at the base of the accreted envelope, upon its mass reaching some critical value \citep{Fujimoto-1982a, Fujimoto-1982b}, and are  usually associated with mass-loss from the system. Nova explosions are relatively common and are a convenient observatory for theoretical and observational study of various aspects of astrophysics, such as binary evolution, theory of accretion onto a compact object, thermonuclear burning on the WD surface and enrichment of the interstellar medium. 

After the optical outburst, these objects are equally interesting to study through their supersoft X-ray emission during their retreat to quiescence. The post-outburst soft X-ray emission results from quiescent nuclear burning of the remnant hydrogen envelope on the WD at a nearly constant bolometric luminosity. This leads to the peak of the spectrum shifting to shorter wavelengths as the photosphere of the WD contracts,  following the expansion of the envelope ejected during the nova (\citealt{Starrfield-3, Sala, Wolf}). The soft X-ray emission also occurs during the peak of the TNR, but only for a very brief period that makes its detection more difficult  (see \citealt{Krautter} and references therein). The post-outburst emission, however, lasts until the remnant envelope is nearly exhausted.  Observations of the post-nova supersoft X-ray emission  provide a means to verify the nova evolution models and to directly probe  parameters of the   WD and the nova explosion. For example, the duration of the post-outburst SSS phase sheds light on the WD mass and the mass ejected during the nova explosion, as was shown by \citet{Tang-2014} and \citet{Henze-2015} for the recurrent nova M31N~2008-12a in M31. Such observational constraints will help fill the missing gaps in   theoretical modelling  of classical and recurrent novae (e.g., \citealt{Starrfield-2, Starrfield-2b, Prialnik-1986, Prialnik-1995, Jose}; see also \citet{Bode} for a review on various aspects of novae).

On the theoretical front of the post-nova evolution study, \citet{Tuchman} invoked the analogy of this phase with the post AGB phase of stellar evolution and derived a modified core mass -- luminosity relation that accounted for the post-nova metal enrichment in the remnant envelope. They applied this relation to observed novae and found good agreement with the observed SSS duration of GQ Mus and V1974 Cyg. Further, \citet{Sala} 
modelled the post-nova evolution for various WD masses by constructing grids of stable hydrogen burning WD envelopes for four different chemical compositions. The photospheric properties of the post-nova SSSs derived from observations, in particular the evolution of their effective temperatures, were then compared with the models to constrain the WD mass, and the envelope mass and composition. These models were applied in the above manner to V1974 Cyg by \citet{Sala-b} and the WD properties for V1974~Cyg were determined. 

Most recently \citet{Wolf} have performed multicycle evolutionary calculations of the post-outburst phase of novae using the MESA (Modules for Experiments in Stellar Astrophysics) code \citep{Paxton-2011, Paxton-2013}. Their calculations covered a  grid of WD masses from $0.6~M_\odot$ to $1.34~M_\odot$, accreting solar composition material. Given that the mass-loss mechanism in novae is still an unresolved issue, for their calculations \citet{Wolf} employed two mass-loss prescriptions, viz., super Eddington winds (SEW) and Roche lobe overflow (RLOF). In these prescriptions, the WD undergoes mass-loss until its photospheric luminosity ($L$) becomes less than an effective local Eddington luminosity ($L_{\rm Edd}$) in the SEW case, or until its radius ($R$) becomes less than the Roche lobe radius ($R_{\rm RL}$) in the RLOF case (see \citet{Wolf} for more details). Their results were found to agree with the measurements of effective blackbody temperature ($T_{\rm eff}$) and turn-off time of the soft X-ray emission from novae observed in M31 \citep{Henze-2, Henze-2013} as well as Galactic sources (see references in \citealt{Wolf}). The ejecta masses predicted by their models were also consistent with the ones derived by \citet{Henze-2} for the observed novae in M31 based on their soft X-ray emission turn-on time. Moreover, \citet{Tang-2014} used the results of \citet{Wolf} to show that the supersoft phase of M31N~2008-12a was most consistent with nuclear burning on the surface of a WD in the mass range of 1.32 -- 1.36~$M_\odot$, a more precise result than the lower bound provided by the recurrence time alone.

On the observational front, much effort has been made following the work of \citet{Pietsch-2005, Pietsch-2007}, which established the prominence of novae as the major class of SSSs in M31. Since then dedicated X-ray monitoring campaigns have been carried out with \textit{XMM-Newton} and \textit{Chandra} in order to search for X-ray counterparts of optically observed novae in the central region of M31. The results from these observations have been published by \citet{Henze-1, Henze-2, Henze-3}. The total number of such novae with detected X-ray counterparts stands now at 79 and 51 of them have their spectra classified as supersoft \citep{Henze-3}. These observational results offer an unprecedented opportunity to study the nova population post-eruption. 

In this paper, we predict the number of post-nova SSSs in M31 theoretically and compare our predictions  with observations. To this end, we use an extended set of multicycle post-nova evolutionary tracks from the calculations of \citet{Wolf}, covering a much finer grid of the WD masses than in the original publication,  ranging from $0.6~M_\odot$ to $1.36~M_\odot$.
For the requisite WD mass distribution in novae, we   derive four different forms, two based on observed optical nova statistics in M31 and two on simple theoretical considerations. With these two ingredients,  we predict the population of post-nova SSSs in M31 and compute their  luminosity function and the $T_{\rm eff}$ distribution.  We also predict the number of these sources to be detected in the the dedicated \textit{XMM-Newton} monitoring of M31 from \citet{Henze-1, Henze-2}, and compare it with the observed number.

The paper is organised as follows: in Sec.~\ref{sec:mass_dist}, we estimate the WD mass distribution in novae in M31. The post-outburst models of novae are presented in Sec.~\ref{sec:pn_SSS}, and in Sec.~\ref{sec:fn} we derive the luminosity function and $T_{\rm eff}$ distribution of the post-nova SSSs in M31 and compare our results with observations in Sec.~\ref{sec:obs_com}. This is followed by a discussion of the uncertainties of our calculations and the magnitude of their effects on our results in Sec.~\ref{sec:discuss}. We finally summarize and conclude in Sec.~\ref{sec:conclude}.

\begin{table*}
\caption{Models of the WD mass distribution in novae}
\label{tab:mass_dist_model}
\renewcommand\arraystretch{1.5}
\begin{tabularx}{\textwidth}{lX}
\hline
Model ID	&Description\\
\hline
M-10	&Distribution obtained using the $t_{2}$ distribution of optically observed novae in M31 (Fig.~\ref{fig:arp_darnley_rate}) and relation between $t_{2}$ and $M_{\rm WD}$ from \citet{Yaron}, assuming WD core temperature of $10^{7}$~K\\
M-30	&Similar to the model M-10, but assuming the WD core temperature to be $3\times 10^{7}$~K for Yaron et al. nova models\\
M-flat	&Flat distribution\\
M-TL	&Distribution from \citet{Truran}\\
\hline
\end{tabularx}
\end{table*}

\section{WD mass distribution in novae in M31}\label{sec:mass_dist}
One of the direct observables for a nova is its decline time, generally reported as the time to decline by 2 or 3~mag ($t_2$ or $t_3$, respectively) from the observed visual peak. In the 1-D nova theory,  this  property of the nova is determined by the WD mass ($M_\textrm{{WD}}$), its core temperature ($T_{\rm c}$) and the mass accretion rate ($\dot{M}$) (see \citealt{Prialnik-1995} and references therein). In real novae, the light curve shape is further affected by possible asymmetry of the explosion and by the interaction of the nova ejecta with the accretion disk and the donor star. These will also introduce orientation-dependent effects, among other consequences. These complications are not included in the currently existing  multicycle nova  models, and therefore accurate  visual decline times for theoretical models are not available. However, it has been shown that the  mass-loss timescale $t_\textrm{{ml}}$ computed in the \citet{Prialnik-1995} and \citet{Yaron} models approximates rather well the  $t_{3}$ time in observed novae \citep{Prialnik-1995}. We will therefore use $t_{\rm ml}$ as a proxy to $t_{3}$ in our calculations.
We can then  use the nova simulation results, giving $t_3\approx t_{\rm ml}(M_{\rm WD}, T_{\rm c}, \dot{M})$,  to approximately map the observed decline time distribution of novae to their WD mass distribution. To do this, we will also have to  make   some assumptions about their mass accretion rates and the WD core temperatures, as discussed below.

Based on the observed orbital period distribution of novae, \citet{Townsley-2005} found that majority of the novae accrete at rate $10^{-9}~M_{\odot}{\rm yr}^{-1}$ driven by magnetic braking. Further, \citet{Townsley-2004} studied the thermal effects of accretion on the WD undergoing nova outbursts and calculated self-consistently an equilibrium core temperature ($T_\textrm{{c,eq}}$) for the WD. This $T_\textrm{{c,eq}}$ is dictated mainly by the mass accretion rate and  by the evolutionary timescale (also determined by $\dot{M}$); the latter characterising whether the WD core has reached the thermal equilibrium. For $\dot{M}=10^{-9}~M_{\odot}{\rm yr}^{-1}$, the WD core should be expected to  nearly achieve the equilibrium  temperature of $T_\textrm{{c,eq}}\approx 10^7$~K. The extended nova evolutionary tracks of \citet{Wolf} have been calculated for $\dot{M}=10^{-9}~M_{\odot}\textrm{yr}^{-1}$ and $T_\textrm{{c}}=3\times 10^7$~K for all except the $1.36~M_\odot$ model where $T_\textrm{{c}}=6\times 10^7$~K was used, i.e., for a probably somewhat higher temperature than expected for the WD in this case. However, this difference is unimportant, as for the post-nova phase, the principal parameter is the envelope mass, and from \citet{Yaron} results for $\dot{M}=10^{-9}~M_{\odot}\textrm{yr}^{-1}$ we see that the nova ignition masses in their Table 2 do not significantly depend on the  core temperature in the considered range of temperatures. 

\begin{figure}
\includegraphics[width=84mm]{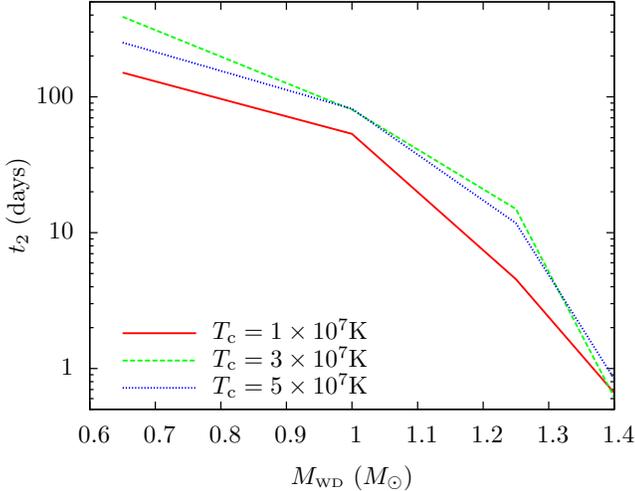}
\caption{ Theoretical decline time $t_2$ of novae as a function of the WD mass from \citet{Yaron} for mass accretion rate $\dot{M}=10^{-9}~M_{\odot}~\textrm{yr}^{-1}$ and different WD core temperatures as indicated in the legend.}\label{fig:t2} 
\end{figure}

In order to calculate the WD mass distribution in novae, we then obtain the decline time as a function of $M_\textrm{{WD}}$ for $\dot{M}=10^{-9}M_{\odot}\textrm{yr}^{-1}$ from \citet{Yaron} by log-linearly interpolating their results between the grid values of the WD mass. For observed extragalactic novae $t_2$ is generally reported since it is easier to measure than $t_3$. The $t_2$ time for the theoretical models from \citet{Yaron} is approximated by $t_{\rm ml}/2.1$ for $t_{\rm ml}<50$~days and by $t_{\rm ml}/1.75$ for $t_{\rm ml}\ge 50$~days following \citet{Bode}. In Fig.~\ref{fig:t2}, this decline time $t_2$ is shown as a function of $M_{\rm WD}$ for the three core temperatures Yaron et al. have used, i.e. $10^7$, $3\times 10^{7}$ and $5\times 10^{7}$~K. As is evident from the figure, the decline times of the novae for $T_{\rm c}=3\times 10^7$ and $5\times 10^{7}$~K agree within a factor of $\lesssim 2$. We will therefore only derive the WD mass distribution for two values of the temperature $T_{\rm c}=10^7$ and $3\times 10^{7}$~K.
As mentioned above, the post-nova evolution should not depend strongly on the core temperature, hence we use the same post-nova models for both $T_{\rm c}$'s in our calculations. For the rest of the paper we adopt the mass accretion rate of the novae to be $10^{-9}~M_{\odot}{\rm yr}^{-1}$.

\setcounter{figure}{1} 
\begin{figure*}
\includegraphics[width=84mm]{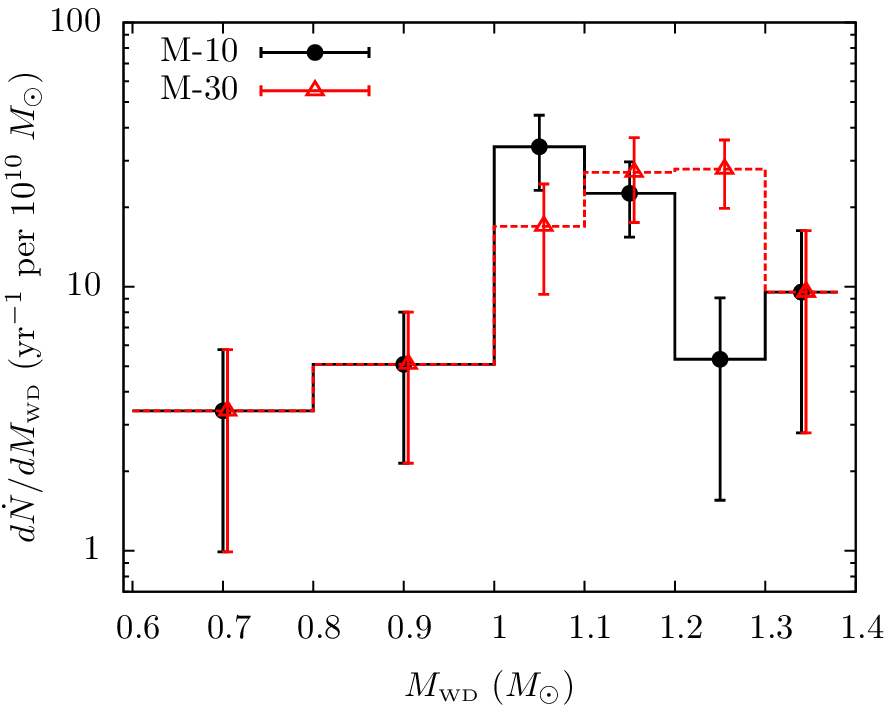}\hfill\includegraphics[width=84mm]{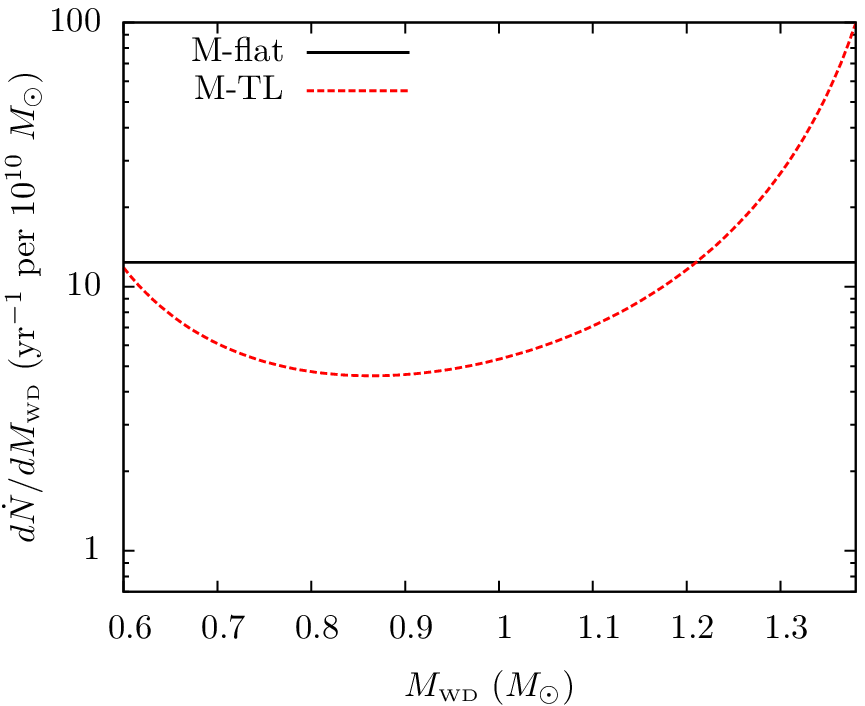}
\caption{ \textit{Left panel}: The inferred WD mass distribution in novae in M31 normalized to $10^{10}~M_\odot$, using the nova samples of \citet{Arp} and \citet{Darnley-2} (cf. Appendices~\ref{append:Arp} and ~\ref{append:rate_dist}). The WD masses of the novae are derived based on their observed decline times (Fig.~\ref{fig:arp_darnley_rate}) using the results from \citet{Yaron} (see text for details). The corresponding mass distribution for WD $T_\textrm{c}=10^7$~K (model M-10 in Table~\ref{tab:mass_dist_model}) is shown here in black, and that for $T_\textrm{{c}}=3\times 10^7$~K (model M-30) in red. The error bars have been derived via error propagation, assuming Poissonian error for the detected number of novae. \textit{Right panel}: Assumed theoretical  WD mass distributions in novae per $10^{10}~M_\odot$ in M31. The black line shows the uniform distribution (model M-flat in Table~\ref{tab:mass_dist_model}) and the red dotted line represents the distribution from \citet{Truran} (model M-TL).}\label{fig:mass_fn}
\end{figure*}

We now use the observed $t_{2}$ distribution of the nova rate in M31 corrected for incompleteness (Fig.~\ref{fig:arp_darnley_rate}), which is described in the Appendices~\ref{append:Arp} and ~\ref{append:rate_dist}, with the results from \citet{Yaron} (Fig.~\ref{fig:t2}) to deduce the corresponding WD mass distribution in novae. The result is shown in Fig.~\ref{fig:mass_fn} (left panel). As one can see from the plot, the two mass distributions are qualitatively similar, except that  for the higher WD temperature,  its peak is shifted towards   higher WD masses by about one mass bin.

In addition, we also explore two simple forms of WD mass distribution in novae, which are independent of the nova models from \citet{Yaron}. Specifically, we consider the distribution from \citet{Truran} as well as an ad hoc flat mass distribution. 
To derive  the former, \citet{Truran}  assumed the \citet{Salpeter} mass function for the WD progenitors and used the main-sequence mass -- WD mass relation for single stars to obtain the distribution of WD masses. Furthermore, they  assumed that the nova recurrence frequency $\nu$  is a function of only the WD mass, and using a simplified treatment of the nova explosion mechanism derived the $\nu=f(M_{\rm WD}$). Finally, they determined the WD mass distribution in novae as a product of $\nu=f(M_{\rm WD}$) and the WD mass function. Although this model is obviously oversimplified, it is not affected by the complexities of the  multicycle nova evolution models and by the subtleties of the incompleteness correction of observed distributions. For this reason we used it, along with an ad hoc flat distribution,  to investigate  the dependence  of our results on the assumed WD mass distribution.

We normalise both the flat mass distribution and that of \citet{Truran}   to the total nova rate in M31 of $106~{\rm yr}^{-1}$, same as our experimentally derived mass distribution (cf. Appendix~\ref{append:rate_dist}). They are shown in Fig.~\ref{fig:mass_fn} (right panel) and their properties are   summarised  in Table~\ref{tab:mass_dist_model}.  


\section{Post-outburst SSS phase of Novae}\label{sec:pn_SSS}

To explore the post-nova SSS phase, we use the nova evolutionary tracks from \citet{Wolf} recomputed on an extended grid of WD masses. These tracks were computed with MESA \citep{Paxton-2011, Paxton-2013} for WD masses 0.6, 0.7, 0.8, 0.9, 1.00, 1.05, 1.10, 1.15, 1.20, 1.25, 1.30, 1.32, 1.34 and 1.36~$M_\odot$ with SEW mass-loss prescription. We will discuss our choice of the  mass-loss prescription towards the end of this section. Using these tracks, we derive the evolution of the nova in the soft X-ray band 0.2--1.0~keV -- the energy range used by \citet{Henze-1, Henze-2, Henze-3} in their analysis of {\it XMM-Newton} data of X-ray monitoring of novae in M31. In computing the soft X-ray light curves of the post-nova SSSs we assume a blackbody emission spectrum. The impact of this assumption on our conclusions is further discussed in Sec.~\ref{sec:discuss}. The photospheric radius and bolometric luminosity are self-consistently computed at each time step of the nova tracks, from which the effective temperature is calculated using the Stefan-Boltzmann law (see \citealt{Wolf} for details). With these quantities given, the 0.2--1.0~keV band luminosity can be straightforwardly computed for the assumed blackbody spectral energy distribution. Typical fractions of the bolometric luminosity emitted in the 0.2--1.0~keV band range from 0.03 for the lowest WD mass to 0.85 for the most massive WDs, i.e., under these assumptions, the bolometric correction for the X-ray band drops to nearly unity as the WD mass approaches the Chandrasekhar mass limit.

The (unabsorbed) soft X-ray light curves are shown in Fig.~\ref{fig:lcs}~(left), after aligning them with respect to the outburst time. Also shown in the right panel are the peak luminosities of these light curves, both unabsorbed as well as the absorbed ones. The latter is obtained by computing the light curves taking into account the Galactic foreground absorption, by assuming a hydrogen column density $N_\textrm{{H}}\approx 6.7\times 10^{20}~\textrm{cm}^{-2}$ towards M31 \citep{Stark} and using the Tuebingen-Boulder absorption model \citep{Wilms}. There is also a SSS phase during the TNR itself, but it lasts at most $\sim1$~day for the luminosity range shown in Fig.~\ref{fig:lcs}, and is not plotted there. However, in our calculations we use the  complete light curve. Shown in Fig.~\ref{fig:T_peak} is the corresponding evolution of the effective temperature $T_{\rm eff}$. 

Since the post-nova SSS phase results from the stable burning of the remnant hydrogen on the WD surface, we expect from nova theory (e.g. \citealt{Prialnik-1995}; \citealt{Yaron}) that more massive WDs, which require smaller ignition mass, will have a shorter SSS phase duration. This is evident in the light curves shown in Fig.~\ref{fig:lcs}. Also, since this phase of the nova is phenomenologically similar to the post-AGB phase of stellar evolution (see \citealt{Sala} and references therein), we expect from the core mass-luminosity relation the more massive WDs with their higher surface gravity to have higher luminosity and accordingly higher effective temperature than the less massive ones. This is  demonstrated  in the right panels of Figs.~\ref{fig:lcs} and \ref{fig:T_peak}.

\setcounter{figure}{2} 
\begin{figure*}
\includegraphics[width=84mm]{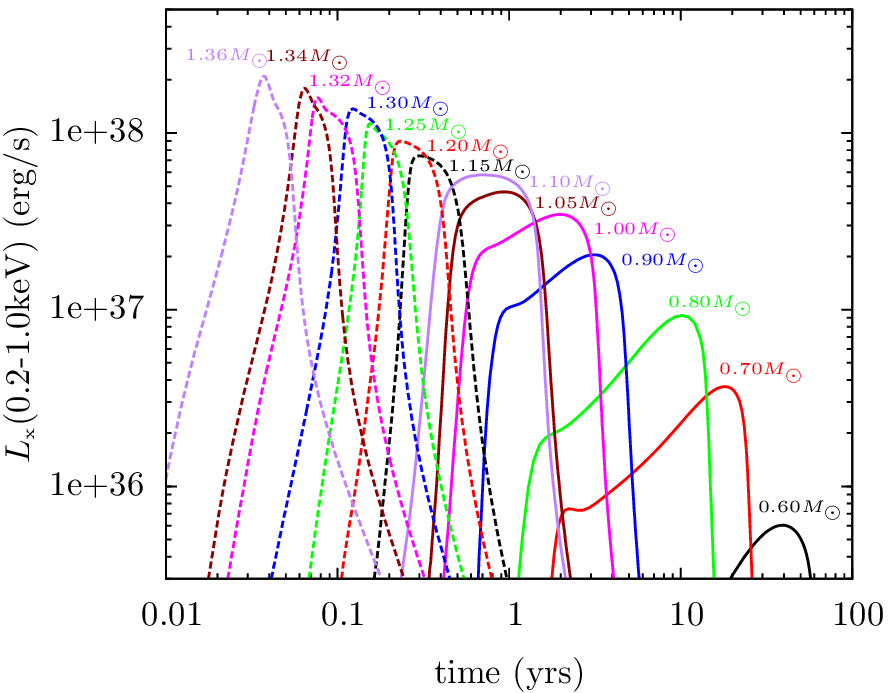}\hfill\includegraphics[width=84mm]{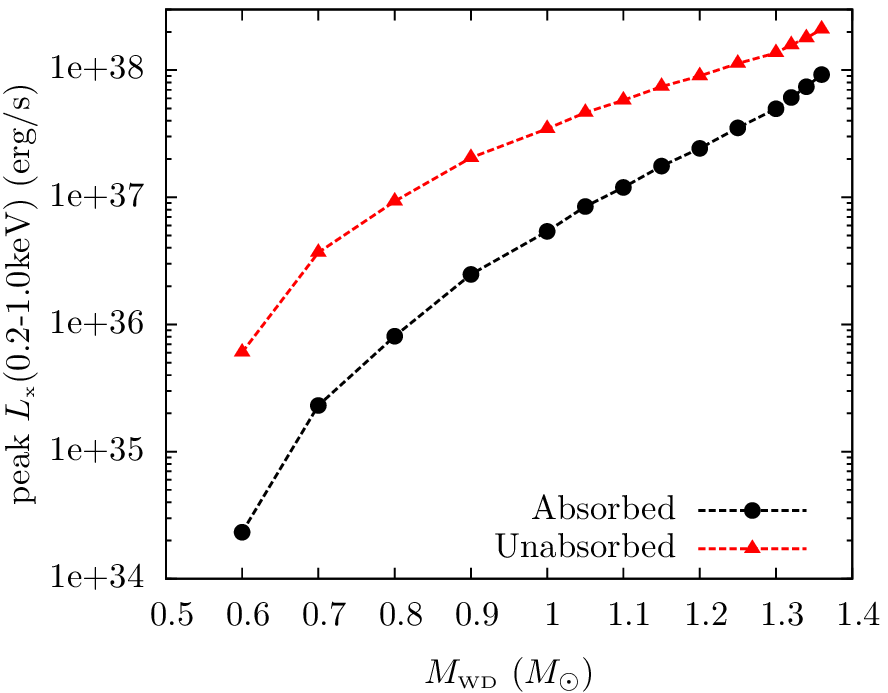}
\caption{\textit{Left panel}: Evolution of the post-nova (unabsorbed) soft X-ray luminosity in the 0.2-1.0~keV band for different WD masses, as indicated against the corresponding curve. These light curves have been aligned with respect to the outburst time. 
\textit{Right panel}: Peak soft X-ray luminosity from the light curves in the left panel as a function of the WD mass. The black circles represent these luminosities with Galactic foreground absorption towards M31 ($N_\textrm{{H}}\approx 6.7\times 10^{20}~\textrm{cm}^{-2}$) and the corresponding unabsorbed luminosities are shown as red triangles.}\label{fig:lcs}
\end{figure*}

\setcounter{figure}{3}
\begin{figure*}
\includegraphics[width=84mm]{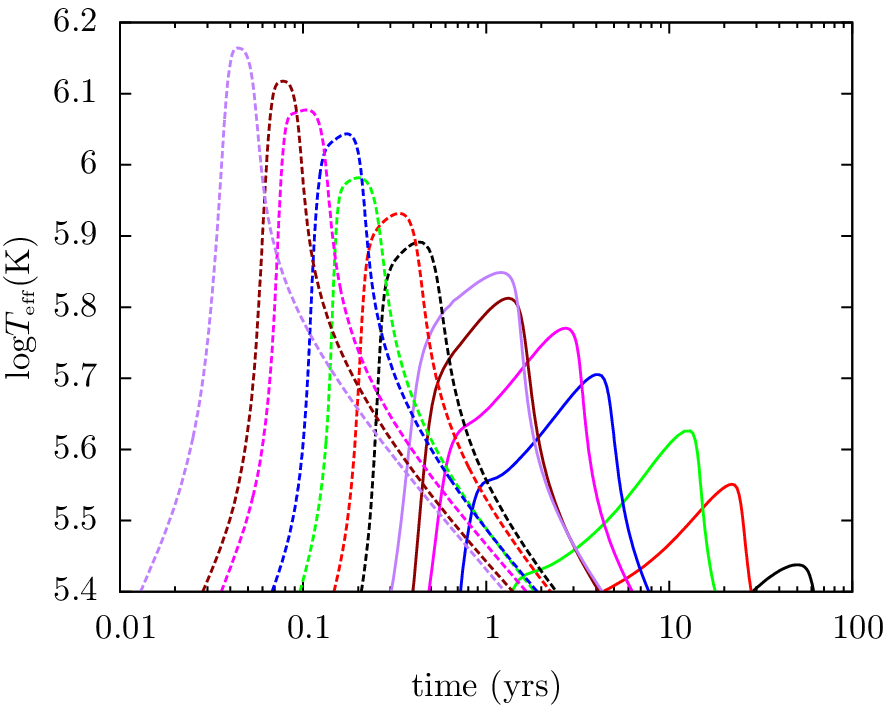}\hfill\includegraphics[width=84mm]{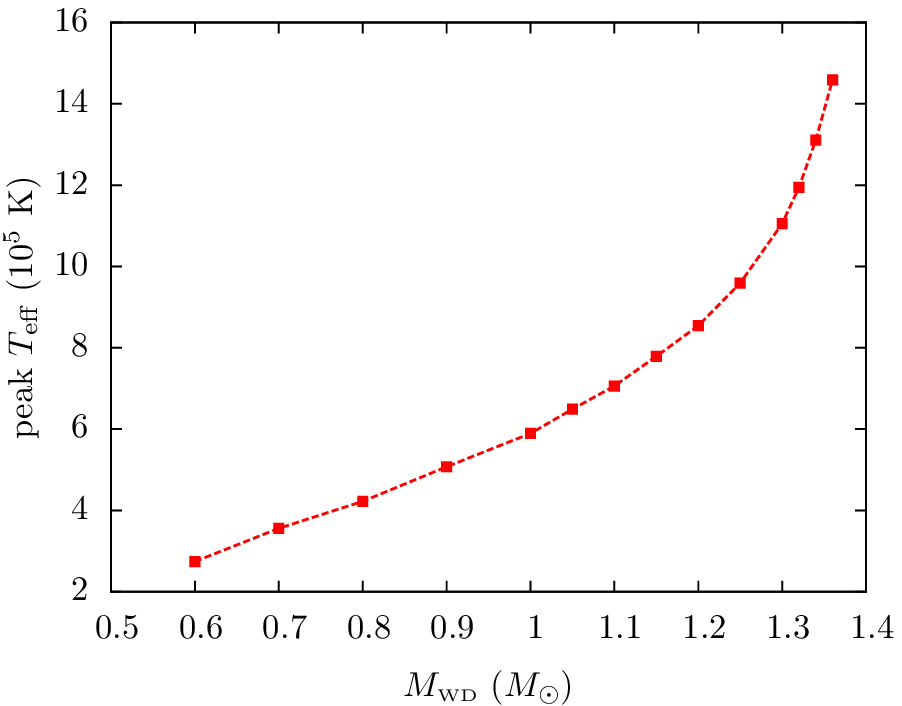}
\caption{ \textit{Left panel}: Evolution of the effective temperature ($T_{\rm eff}$) of post-nova SSSs from the nova evolutionary tracks of \citet{Wolf} computed using the Stefan-Boltzmann law (see \citet{Wolf} for details). The colour coding is the same as in Fig.~\ref{fig:lcs}. These $T_{\rm eff}$ curves have also been aligned with respect to the outburst time. \textit{Right panel}: Peak effective temperature as a function of the WD mass.}\label{fig:T_peak}
\end{figure*}

As mentioned earlier, we use the post-nova evolutionary models from \citet{Wolf}, which employ the SEW mass-loss mechanism. 
It is known that the mass-loss process in novae is still not well understood. In fact, for novae with short recurrence period occurring on massive WDs, the dominant uncertainty in their modeling is probably the mass-loss prescription as compared to other uncertainties, such as mixing (see Sec.~\ref{sec:discuss}). In nova models, besides the SEW and RLOF mass-loss prescriptions (Sec.~\ref{sec:intro}) used by \citet{Wolf}, there are other prescriptions used by different groups that may or may not be more valid, such as the optically thick wind used by \citet{Kato-1994} or the hydrodynamic outburst followed by wind implemented by \citet{Prialnik-1995} and \citet{Yaron}. The main effect of these different wind prescriptions on the post-nova SSS phase is that, for the same amount of ejected material, the faster winds will make the ejecta optically thin faster, revealing the SSS earlier. Alternatively, for the same mass-loss duration, some prescriptions will eject more material than the others, such as in the case of SEW versus RLOF \citep{Wolf}.

From the power law fits of \citet{Wolf} to the  dependence of the turn-off time of post-nova SSS phase on the WD mass, the duration of this phase for the RLOF models is $\sim 4$ times that of the SEW models. This could have a significant impact on the  results of our calculations. However, according to \citet{Wolf}, the SEW models are more appropriate for novae occurring on massive WDs while the RLOF models are suitable for less massive ones. Given the typical sensitivity of the current X-ray observations, including those used for comparison later in this paper, we should expect to observe mainly the bright post-nova SSSs, occurring on sufficiently massive WDs, $\ga 1~M_\odot$ (Fig.~\ref{fig:lcs}). Therefore, between SEW and RLOF models, the use of those with SEW mass-loss prescription is justified.

\section{Post-nova SSS population in M31}\label{sec:fn}

\begin{table}
\caption{Theoretical nova rate for the different WD mass distribution models. Masses are in solar units and for each model, rates are given per year in the given WD mass interval range. }
\label{tab:rate}
\renewcommand\arraystretch{1.1}
\begin{tabular}{lcllll}
\hline
mass range	&mid-mass	&M-10	&M-30	&M-flat	&M-TL\\
\hline
0.650 - 0.750	&0.70	&3.72	&3.72	&13.59	&6.85\\
0.750 - 0.850	&0.80	&4.66	&4.66	&13.59	&5.30\\
0.850 - 0.950	&0.90	&5.59	&5.59	&13.59	&5.15\\
0.950 - 1.025	&1.00	&12.10	&7.45	&10.19	&4.30\\
1.025 - 1.075	&1.05	&18.62	&9.31	&6.79	&3.32\\
1.075 - 1.125 	&1.10	&15.51	&12.10	&6.79	&3.92\\
1.125 - 1.175	&1.15	&12.40	&14.90	&6.79	&4.87\\
1.175 - 1.225	&1.20	&7.66	&15.11	&6.79	&6.43\\
1.225 - 1.275	&1.25	&2.92	&15.32	&6.79	&9.24\\
1.275 - 1.310	&1.30	&2.51	&8.71	&4.76	&9.62\\
1.310 - 1.330	&1.32	&2.10	&2.10	&2.72	&7.59\\
1.330 - 1.350	&1.34	&2.10	&2.10	&2.72	&10.12\\
1.350 - 1.370	&1.36	&2.10	&2.10	&2.72	&14.33\\

\hline
\end{tabular} 
\end{table}

With the WD mass distribution in novae in M31 (Sec.~\ref{sec:mass_dist}) and the post-nova models (Sec.~\ref{sec:pn_SSS}), we can compute the post-nova SSS population. 
Their  luminosity function is given by
\begin{equation}\label{eq:diff}
\dfrac{dN(L_\textrm{{x}})}{dL_\textrm{{x}}}=\int^{M_\textrm{ch}}_{M_\textrm{low}}\dfrac{d\dot{N}(M_{\textrm{{WD}}})}{dM_{\textrm{{WD}}}}\phi(L_{\rm x},M_{\textrm{{WD}}}) dM_{\textrm{{WD}}},
\end{equation}
where $M_{\rm ch}$ is the Chandrasekhar mass limit, $M_{\rm low}$ is the lower limit of  the WD mass. The first term inside the integral on the right hand side is the WD mass distribution in novae and the function $\phi(L_{\rm x},M_{\textrm{{WD}}})$ depends on the shape of the light curve as follows: 
\begin{align*}
\phi(L_{\rm x},M_{\textrm{{WD}}}) &=\left(\dfrac{dL_{\rm x}}{dt}\right)^{-1}_{\rm rise}+\left(\dfrac{dL_{\rm x}}{dt}\right)^{-1}_{\rm decay} &, L_{\rm x}<L_{\rm p}\\
  &=0 &, L_{\rm x}\geq L_{\rm p}\\
\end{align*}
where $L_{\rm x}=L_{\rm x}(t,M_{\rm WD})$  is the (post-nova) soft X-ray  light curve for the WD of mass $M_{\rm WD}$ and $L_{\rm p}=L_{\rm p}(M_{\rm WD})$ is its peak luminosity. The derivatives are taken in the rising and declining parts of the light curve at the points in time when the luminosity equals $L_{\rm x}$.   

As can be seen from Fig.~\ref{fig:lcs} (right panel), the peak (absorbed) soft X-ray luminosity $L_{\rm x}$ for WD masses below $\sim 0.65~M_\odot$ is less than $10^{35}$~erg/s, which is below the  sensitivity limit for observation of M31 by the current X-ray satellites like \textit{Chandra} and \textit{XMM-Newton}. For our calculations we are interested in the bright sources, observable with these X-ray missions. We therefore set $M_{\rm low}=0.65$ in Eq.~\eqref{eq:diff} and solve it using a Monte-Carlo method. 
To this end, we divide the WD mass range $0.65-1.37~M_\odot$ into 13 bins and use the WD mass distribution derived in Sec.~\ref{sec:mass_dist} to obtain the nova rate in each bin as given in Table~\ref{tab:rate}. For each mass bin, we then seed novae randomly in time according to their corresponding rates and follow their luminosity (Fig.~\ref{fig:lcs})  and effective temperature evolution (Fig.~\ref{fig:T_peak}).  To minimise the Monte-Carlo errors, we perform the simulations for a period of $5\times 10^{4}$~yrs, taking 1000 snapshots separated by 50~yrs, a long enough interval for (absorbed) luminosity greater than $10^{35}$~erg/s to ensure that any post-nova SSS is captured in only one snapshot. In each snapshot, we register the number of SSSs with their luminosities and temperatures. Averaging over the snapshots, we calculate the instantaneous number of post-nova  SSSs in M31 as a function of (unabsorbed) luminosity and $T_{\rm eff}$. The results are discussed in the next two sections.

\subsection{Luminosity function}\label{sec:lum_fn}

\setcounter{figure}{4}
\begin{figure*}
\includegraphics[width=84mm]{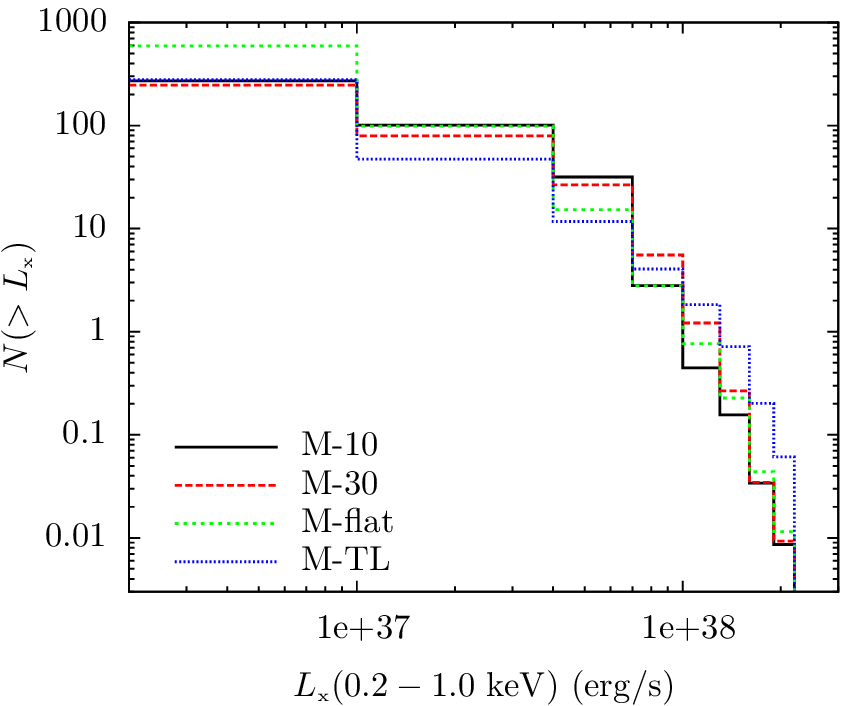}\hfill\includegraphics[width=84mm]{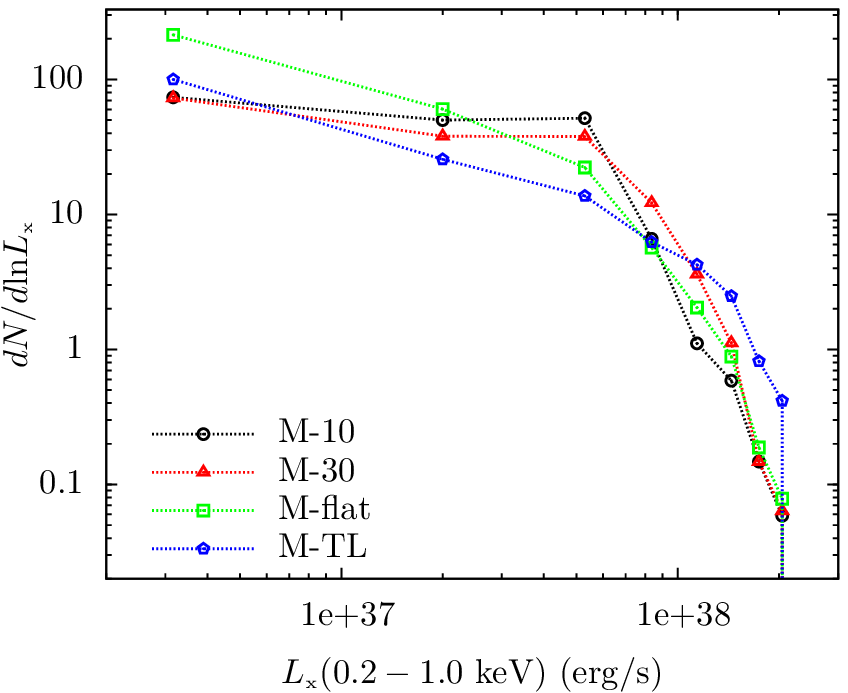}
\caption{ Cumulative (\textit{left}) and differential (\textit{right}) luminosity functions of the post-nova SSSs in M31 for the different WD mass distribution models.  
Note that the luminosity, $L_\textrm{{x}}$, shown in both plots is the unabsorbed luminosity. These functions are normalized to the total stellar mass of M31, which is $\approx 1.1\times 10^{11}~M_\odot$. The Monte-Carlo errors are negligibly small and are not plotted.}\label{fig:lum}
\end{figure*}

\setcounter{figure}{5} 
\begin{figure*}
\includegraphics[width=84mm]{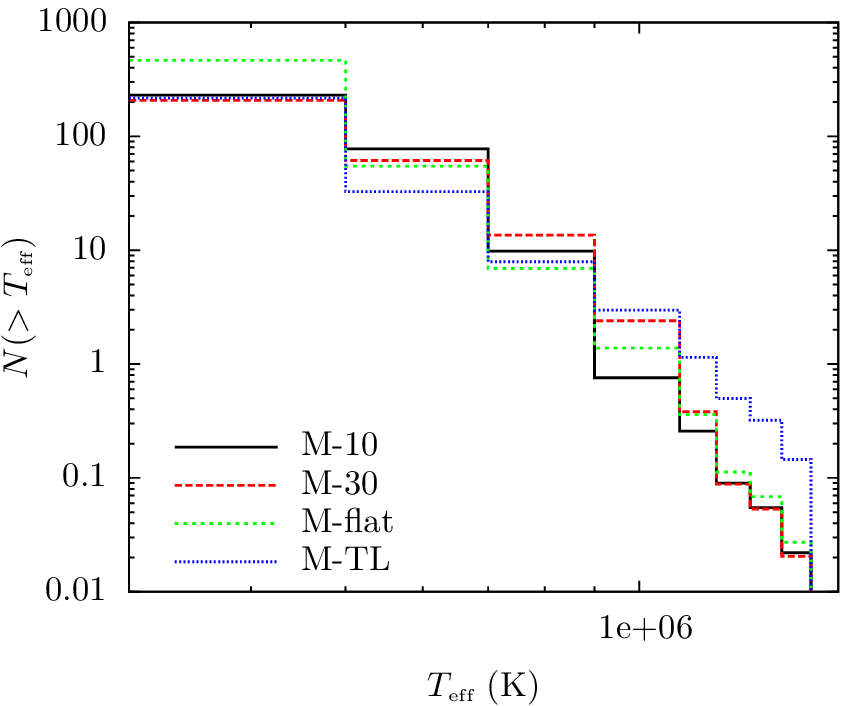}\hfill\includegraphics[width=84mm]{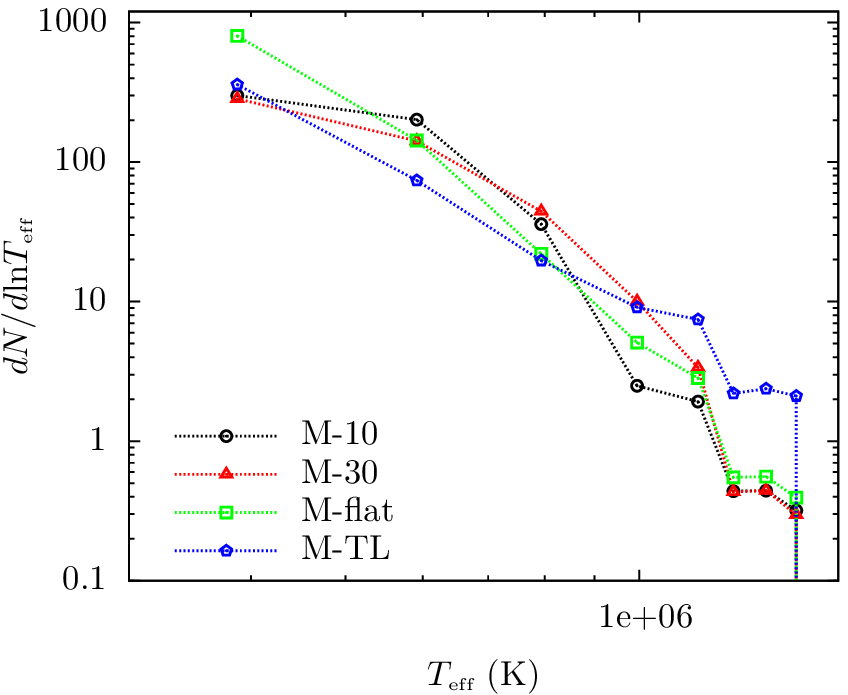}
\caption{Cumulative effective temperature ($T_\textrm{{eff}}$) distribution of post-nova SSSs in M31 with unabsorbed luminosity $L_{\rm x}\gtrsim 10^{36}$~erg/s (\textit{left}) and the corresponding differential distribution (\textit{right}). These distributions are normalized to the total stellar mass of M31, $1.1\times 10^{11}~M_\odot$.}\label{fig:T}
\end{figure*}

The luminosity functions (LFs) of post-nova SSSs obtained under  different assumptions about  the WD mass distribution are shown in Fig.~\ref{fig:lum}. The luminosity bin width is determined by the WD mass sampling of the post-nova evolutionary tracks (cf.~Fig.~\ref{fig:lcs}). We used coarser bins for low luminosities where the tracks are sparser and finer bins for high luminosities where the tracks sample the WD mass with smaller step. Note that the functions are plotted against the unabsorbed luminosity, computed for the 0.2 -- 1.0~keV band. The effects of interstellar absorption in M31 will be considered in Sec.~\ref{sec:obs_com}.

As can be seen from Fig.~\ref{fig:lum}, the LFs in all four cases have a cut-off at $L_{\rm x}\approx 2\times 10^{38}$~erg/s, which corresponds to the maximum unabsorbed soft X-ray luminosity achieved by the post-nova X-ray sources containing the most massive WDs (see Fig.~\ref{fig:lcs} right). Accounting for bolometric correction, this cut-off luminosity exceeds marginally the Eddington limit for the most massive WDs ($1.36~M_{\odot}$) in the grid of post-nova models of \citet{Wolf} used in our calculations. The occurrence of luminosity $L_{\rm x}$ close to the Eddington limit in our results can be understood as follows. The post-nova models of \citet{Wolf} used in our calculations do not avoid super Eddington luminosities. In fact, as discussed in Sec.~\ref{sec:pn_SSS}, the mass loss during the nova outburst is regulated by a wind triggered by the luminosity in excess of the Eddington limit (see \citet{Wolf} and \citet{Denissenkov-2013} for details). This in turn acts to reduce the nuclear burning, which is driven by the pressure of the envelope acting at the base of the hydrogen layer, causing the luminosity to become sub-Eddington sooner than it would have in the absence of the wind. However, since the most massive WDs become transparent to soft X-rays sooner than the lower mass WDs (see Fig.~\ref{fig:lcs} left), the occurrence of super Eddington luminosity during the post-nova SSS phase is possible for the former WDs. Also, as discussed in Sec.~\ref{sec:pn_SSS}, the bolometric correction for the soft X-ray luminosity of the massive WDs is smaller than that of the lower mass WDs. Therefore, the maximum $L_\textrm{{x}}$ during the SSS phase for the post-nova models of massive WDs tend to be close to their Eddington limit (cf.~Fig.~\ref{fig:lcs} right).

From the cumulative luminosity function (Fig.~\ref{fig:lum} left), we compute the number of post-nova SSSs with (unabsorbed) $L_{\rm x}\gtrsim 10^{36}$~erg/s at any given time in M31 for the various WD mass distribution models. These numbers, denoted as $n(L_{\rm x}\gtrsim10^{36}~{\rm erg/s})$, are given in Table~\ref{tab:model_results}. We determine the contribution of bright sources to the computed values of $n(L_{\rm x}\gtrsim10^{36}~{\rm erg/s})$ for the different mass distribution models. For model M-10, we find $7\%$ of these sources have $L_{\rm x}\gtrsim5\times10^{37}$~erg/s and $37\%$ have $L_{\rm x}\gtrsim10^{37}$~erg/s. The corresponding fractions for model M-30 are $7\%$ and $32\%$, for model M-flat they are $2\%$ and $17\%$ and for model M-TL they are $3\%$ and $17\%$. Finally,  using the differential LFs, shown in Fig.~\ref{fig:lum} (right panel), we compute,  for each WD mass distribution model,  the combined  unabsorbed luminosity of sources with   $L_{\rm x}\gtrsim10^{36}~{\rm erg/s}$. These results are also listed  in Table~\ref{tab:model_results}.

From Eq.~\eqref{eq:diff} it is clear that the general shape of the differential LFs is determined by low mass and high mass WDs at the low and high luminosity ends, respectively (cf. Fig.~\ref{fig:lcs}, right panel). The steep decline at the high luminosity end for the differential LFs is due to the short SSS duration for the massive WDs with higher peak luminosities, whereas the relatively gradual slope towards low luminosities is due to the longer SSS duration for the less massive WDs with lower peak luminosities. The models M-flat and M-TL have larger number of faint sources ($L_{\rm x}\lesssim10^{37}$~erg/s) than models M-10 and M-30, since the WD mass distributions for M-flat and M-TL have greater values than M-10 and M-30 at WD masses $\lesssim 0.8-1.0~M_{\odot}$ (cf.~Fig.~\ref{fig:mass_fn}). On the other hand, for $L_{\rm x}$ in the range $\sim 10^{37}-10^{38}$~erg/s, models M-10 and M-30 have larger number of sources (see above) as the WD mass distributions for these models attain their maximum at the WD mass values, $\sim 1.0-1.2~M_{\odot}$, having peak luminosities in $10^{37}-10^{38}$~erg/s range (Figs.~\ref{fig:mass_fn} and \ref{fig:lcs}). Finally, beyond $1.3~M_{\odot}$, model M-TL has the largest values for the WD mass distribution (Fig.~\ref{fig:mass_fn}) and accordingly it has the largest number of sources above $\sim 10^{38}$~erg/s, where WDs with mass greater than $1.3~M_{\odot}$ have their peak luminosities, as observed for the recurrent nova M31N~2008-12a.

Given the remarkable nature of this M31 nova M31N~2008-12a, which apparently is ``on'' in soft X-rays for about two weeks every year (e.g., see \citealt{Henze-2014, Tang-2014, Henze-2015}), we have used its X-ray observation results to obtain the range of its (blackbody) effective temperature and corresponding luminosity during its SSS phase. For the $T_{\rm eff}$ range $1$ -- $1.3\times 10^{6}$~K and corresponding photospheric radius from \citet{Tang-2014}, we find its (unabsorbed) luminosity in the 0.2--1.0~keV band to be in the range $1.0$ -- $3.0\times 10^{38}$~erg/s during its ``on'' state (see also \citealt{Henze-2015}). From the observed duration of its SSS phase, we infer the number of this system visible at any instant in M31 to be $\approx 0.04$ with $L_{\rm x}\gtrsim 10^{38}$~erg/s and $T_{\rm eff}\gtrsim 10^{6}$~K. 

Note that the LFs (Fig.~\ref{fig:lum}) and $T_{\rm eff}$ distributions (see below; Fig.~\ref{fig:T}) are the predicted results for the whole M31 galaxy, not subjected to the selection biases of any survey. Further, besides M31N~2008-12a, there will also exist some number of novae in M31 hosting massive WDs, even if their recurrence time is not as short as one year of M31N~2008-12a (cf.~Fig.~\ref{fig:mass_fn}). As discussed above, these post-nova SSSs with massive WDs will have high luminosity as well as high $T_{\rm eff}$ (Sec.~\ref{sec:T_fn}). 
The observation results from this single nova M31N~2008-12a then give a lower limit on the cumulative LF of the post-nova SSSs in M31 at $L_{\rm x}\approx 10^{38}$~erg/s (Fig.~\ref{fig:lum} left) and also on their cumulative $T_{\rm eff}$ distribution at $T_{\rm eff}\approx 10^{6}$~K (Fig.~\ref{fig:T} left). To justify this further, in \citet{Henze-1}, for example, one can find other post-nova SSSs in M31 observed with {\em XMM-Newton}, whose $L_{\rm x}$~(0.2--1.0~keV) obtained from blackbody spectral fitting is $\gtrsim 10^{38}$~erg/s.

\subsection{Effective temperature distribution}\label{sec:T_fn}

We also determine the effective temperature distribution of the post-nova SSSs, whose LFs were obtained in the previous section. The resulting cumulative and differential $T_{\rm eff}$ distributions for  sources with (unabsorbed) $L_{\rm x}\gtrsim 10^{36}$~erg/s are shown in Fig.~\ref{fig:T}.

The analytical expression for the differential $T_\textrm{{eff}}$ distribution is similar to that of the differential luminosity function given by Eq.~\eqref{eq:diff}, but with $\phi(L_{\rm x}, M_{\rm WD})$ replaced with $\psi(T_{\rm eff}, M_{\rm WD})$ defined as 

\begin{align*}
\psi(T_{\rm eff},M_{\rm WD}) &=\left(\dfrac{dT_{\rm eff}}{dt}\right)^{-1}_{\rm rise}+\left(\dfrac{dT_{\rm eff}}{dt}\right)^{-1}_{\rm decay} &, T_{\rm eff}<T_{\rm p}\\
  &=0 &, T_{\rm eff}\geq T_{\rm p}\\
\end{align*}
where  $T_{\rm p}=T_{\rm p}(M_{\rm WD})$ is the peak value of the temperature curve  $T_{\rm eff}=T_{\rm eff}(t,M_{\rm WD})$  for  the corresponding WD mass $M_{\rm WD}$.

As can be seen from Fig.~\ref{fig:T} (right), the differential $T_\textrm{{eff}}$ distributions for all the four WD mass distribution models have a cut-off at $T_{\rm eff}\sim 1.5\times 10^{6}$K. This cut-off value corresponds to the highest $T_{\rm eff}$ obtained from the evolutionary tracks (Fig.~\ref{fig:T_peak} right). 
Further, in close analogy with the LFs, at low $T_{\rm eff}$ values the differential distribution is determined by low mass WDs while at the high $T_{\rm eff}$ end, it is determined by the massive WDs (cf. Fig.~\ref{fig:T_peak} right). The short timescale of the $T_{\rm eff}$ evolution for massive WDs (see Fig.~\ref{fig:T_peak}) leads to smaller SSS numbers in the differential $T_{\rm eff}$ distribution at high $T_{\rm eff}$ end, while the longer duration for the less massive ones leads to higher SSS numbers at low $T_{\rm eff}$ values.

\begin{figure}
\includegraphics[width=84mm]{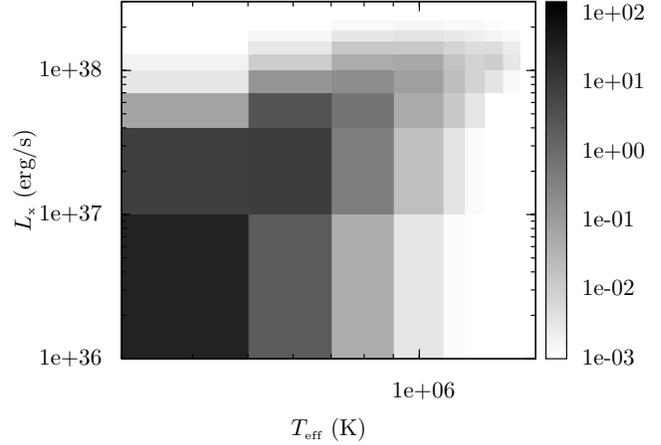}
\caption{ Effective temperature--luminosity plot for the post-nova SSSs in M31 corresponding to the model M-10. The respective plot for the other three models (M-30, M-flat and M-TL) look qualitatively similar. The gray scale indicates the absolute number of SSSs per bin. Note that $L_{\rm x}$ is the integrated unabsorbed luminosity in the 0.2-1.0~keV band.}\label{fig:snap}
\end{figure}

Finally, in Fig.\ref{fig:snap}, we plot the distribution of the post-nova SSSs on the effective temperature -- luminosity plane.  Fig.\ref{fig:snap} shows this distribution for the M-10 model; for other WD mass functions this distribution looks qualitatively similar.  As could be expected, the left bottom corner of the $T_{\rm eff}-L_{\rm x}$ parameter space, i.e. low luminosity and low effective temperature, is relatively more populated.

\begin{table*}
\begin{minipage}{\textwidth}
\caption{Monte-Carlo simulation results for the different models of WD mass distribution in novae in M31}
\label{tab:model_results}
\renewcommand\arraystretch{1.5}
\begin{tabularx}{\textwidth}{cXXXXX}
\hline
Model	&	&$n(L_{\rm x}\gtrsim 10^{36}~{\rm erg/s})${\footnote{Number of post-nova SSSs with unabsorbed $L_{\rm x}\gtrsim 10^{36}$~erg/s at any instant in M31.}}	&$L_{\rm total}~({\rm erg/s})${\footnote{Total luminosity from the $n(L_{\rm x}\gtrsim 10^{36}~{\rm erg/s})$ post-nova SSSs.}}	&$N_{\footnotesize{XMM-Newton}}${\footnote{Number of post-nova SSSs predicted for the \textit{XMM-Newton} observations from \citet{Henze-1, Henze-2}. The first value corresponds to the threshold (unabsorbed) luminosity $L_{\rm x}=5\times 10^{36}$~erg/s and the second value to $L_{\rm x}=3\times 10^{36}$~erg/s (cf. Sec.~\ref{sec:obs_com}).}}	&$\Delta N_{\rm H}~(10^{21}{\rm cm}^{-2})${\footnote{Neutral hydrogen column density of the circumstellar material required to reduce the predicted number of sources for the \textit{XMM-Newton} observations to 16 that were observed (cf. Secs.~\ref{sec:obs_com} and \ref{sec:discuss}). The first value is for threshold (unabsorbed) luminosity $L_{\rm x}=5\times 10^{36}$~erg/s and the second value for $L_{\rm x}=3\times 10^{36}$~erg/s.}}\\
\hline
M-10	&	&271	&${3.97\times 10^{39}}$	&65--74		&1.70--2.35\\
M-30	&	&247	&${3.44\times 10^{39}}$	&57--65		&2.05--2.75\\
M-flat	&	&592	&${4.67\times 10^{39}}$	&48--63		&1.25--1.85\\
M-TL	&	&277	&${2.51\times 10^{39}}$	&33--40		&1.45--2.05\\
\hline
\end{tabularx}
\end{minipage}
\end{table*}

\section{Comparison with \textit{XMM-Newton} observations of M31}\label{sec:obs_com}

Our knowledge of the statistics of novae detected  in the post-outburst SSS phase has improved significantly in the past years, especially thanks to extended XMM-Newton monitoring campaign  of M31. We therefore take this opportunity to compare our theoretical predictions with observational results. In particular, we use the results of \citet{Henze-1, Henze-2}  based on \textit{XMM-Newton} observations of the central region of M31. This monitoring program comprised three campaigns -- June 2006 to March 2007, November 2007 to February 2008 and November 2008 to February 2009. In total there were 15 observations, all pointed at the center of M31, with typical exposure time of around 10-20~ksec each. Details of these observations are given in \citet{Henze-1, Henze-2}. 

In order to account for the limited spatial coverage of M31 in these observations, we normalize the nova rate in our simulations to the mass contained within the field of view (FOV) of \textit{XMM-Newton}. To determine this mass we use the Spitzer 3.6~micron image of M31 from \citet{Barmby} and approximate the {\it XMM-Newton}  FOV by a circular region centered at the nucleus of M31 with a diameter of $30'$. We then compute the stellar mass enclosed in the same manner as in Appendix~\ref{append:rate_dist}. These M31 \textit{XMM-Newton} observations suffer from source confusion in the innermost region \citep{Henze-1, Henze-2, Henze-3}. Following \citet{Henze-3}, we  subtract  from the calculated total stellar mass within the FOV, the stellar mass enclosed in a rectangular region of dimensions $3.3'\times3.3'$ centered at the nucleus of M31. We thus obtain a value of $\approx 3\times 10^{10}~M_\odot$, which is used to normalize the nova rate. 

We repeated the Monte-Carlo simulations described in section \ref{sec:fn} with the  WD mass distribution renormalised according to the nova rate within the {\it XMM-Newton} FOV. However, now we placed the snapshots in time according to the pattern of {\it XMM-Newton} observations.  We obtain a catalogue of the sources appearing in these snapshots, registering their luminosities and effective temperatures.  We then filter this catalogue using the detection sensitivity of \textit{XMM-Newton} (see below) to obtain the actual number of  sources ``detected'' in these simulations. To minimise the Monte-Carlo errors, we repeated the simulations multiple number of times.

According to \citet{Henze-2}, the  detection sensitivity of \textit{XMM-Newton} in a typical  individual observation from the M31 monitoring program is around a few times $10^{36}$~erg/s in the  0.2-1.0~keV band. We therefore ran two sets of simulations with the sensitivity limit of  $3\times 10^{36}$ and $5\times 10^{36}$~erg/s. The quoted limiting luminosity from \citet{Henze-2} is the unabsorbed luminosity of the source derived assuming a 50~eV blackbody spectrum and Galactic foreground absorption. We used PIMMS \footnote{\url{http://heasarc.gsfc.nasa.gov/docs/software/tools/pimms.html}} to recompute it for the given value of the effective temperature (different for different sources ``detected'' in our simulations) and for the  hydrogen column density $N_{\rm H}$ taking into account the spatially varying intrinsic absorption in M31.

To incorporate the latter, we use the HI map from \citet{Brinks} (see also \citealt{Nieten}).  We construct a grid over the \textit{XMM-Newton} FOV with cell size  $2'\times2'$ and use the HI map to calculate the total $N_{\rm H}$ in each cell, including the Galactic foreground value of $6.7\times10^{20}~\textrm{cm}^{-2}$. In each grid cell, for each source temperature from the above catalog of simulated sources we determine the detection threshold and thereby  compute the total number of sources, whose luminosity exceeds the detection threshold. As the source catalogue was derived for the entire XMM-Newton FOV, the number of sources in the given grid cell is scaled down according to the stellar mass contained in the cell.  This procedure is carried out for all the cells and the results are summed up to calculate the total number of ``detected'' sources.  The results of these calculations for various WD mass distribution models are summarised  in Table~\ref{tab:model_results}, from which we can see that according to our calculations,  about ${\approx 30-80}$ post-nova SSSs should be detected in the XMM-Newton monitoring campaign.

\citet{Henze-1, Henze-2} found 16 SSSs in total from these observations, located within the FOV we have considered, 13 of which have been identified with previously registered optical novae. Given the incompleteness of the optical nova surveys, it is not unlikely that the remaining 3 SSSs are also associated with novae. From Table~\ref{tab:model_results}, we thus see that our predictions agree with XMM-Newton observations within  a factor of ${\approx 2-5}$. These results are discussed further in the following section.

\section{Discussion}\label{sec:discuss}

Given the number and magnitude of uncertainties involved in modelling the post-nova supersoft X-ray phase, the factor of ${\approx 2-5}$ disagreement between the predicted number of post-nova SSSs and the result of {\em XMM--Newton} observations is remarkably modest. Moreover, it is comparable to the spread (about a factor of $\sim 2$) between the results obtained for different models of the WD mass distribution in novae. 
Below we discuss how these results depend on various uncertainties and assumptions involved in our calculations.

The main simplification of the calculations presented in this paper is the assumption of blackbody spectrum for the post-nova SSSs. This assumption is commonly made in SSS studies; it is motivated, at least partly,  by the fact that  the observed spectral energy distribution (SED) of the continuum emission in SSSs can be often represented by the simple blackbody spectrum \citep[e.g,][]{Skopal}. However, it is obvious that spectral formation in SSSs is much more complex. Indeed, a number of emission and absorption lines and absorption edges are observed in the spectra taken with better sensitivity and energy resolution, with some evidence that orientation effects may play a significant role \citep{Ness}. Currently, no theoretical model can describe  the entire  complexity of  SSS spectra. However, the  existing  static NLTE WD atmosphere calculations  may serve as a useful first approximation to gauge the magnitude of the effect on our results from using more realistic spectral models. To this end, we used the publicly available TMAP models \citep{Rauch-2010} provided by the TheoSSA service (\url{http://dc.g-vo.org/theossa}).
For the range of interesting WD masses (0.90--1.25~$M_\odot$) and their peak effective temperatures ($5\times 10^{5}$ -- $10^{6}$~K) , we compare the 0.2--1.0~keV luminosity of a blackbody spectrum with that from the NLTE models with the same effective temperature.
We find the NLTE (unabsorbed) luminosities to be lower than the blackbody ones typically by a factor of $\sim2$.   
Because of the rather flat shape of the luminosity distribution of the post-nova SSSs (Fig.~\ref{fig:lum}) in the luminosity range of interest, $\log(L_{\rm x})\sim 36$ -- 37, this does not have a significant effect on the total number of sources. From Fig.~\ref{fig:lum}, we estimate that this will decrease the predicted number of post-nova SSSs in Table~\ref{tab:model_results} by  a factor of $\approx 1.3$. Comparing the {\em absorbed} blackbody and NLTE luminosities, computed using a total $N_{\rm H}$ of $2\times10^{21}~{\rm cm}^{-2}$ from the Galactic foreground ($6.7\times 10^{20}~{\rm cm}^{-2}$) and the interstellar medium in M31 (typically $\sim 10^{21}~{\rm cm}^{-2}$), we now find a factor of 1.2 -- 1.7 difference between the two for the above WD parameters. We therefore conclude that the use of  NLTE models will reduce the discrepancy between our predictions and the \textit{XMM-Newton} observations.

It should be noted however that the stationary TMAP NLTE models do not quite reproduce the entire complexity of the observed SSS spectra, in particular the significant orientation effects and the blue shifts of absorption lines seen in X-ray grating spectra of Galactic post-nova SSSs \citep{Ness-2010, Ness}. The blue shifts indicate the presence of an expanding nova shell \citep{Rossum-2010}, which is not accounted for in TMAP models as they assume plane parallel geometry and hydrostatic equilibrium. This restriction has been lifted in the wind model of \citet{Rossum-2010} and \citet{Rossum-2012}. The latter used the wind models to  fit the observed grating spectra of nova V4743 Sgr 2003 during its SSS phase. They obtained a slightly different $T_{\rm eff}$ than \citet{Rauch-2010b} based on TMAP model. Direct application of the wind models of \citet{Rossum-2012} to our calculations is not straightforward however, as the parameters required for the wind models, specifically the expansion velocity and mass outflow rate, necessitate their consistent integration with the post-nova modeling, which is beyond the scope of this paper. This will be pursued in future
as a follow-up to this work. On the other hand, the results of \citet{Skopal} suggest that the spectral energy distribution of many SSSs is nevertheless broadly consistent with the blackbody spectrum, i.e., the above mentioned effects should not affect the bolometric correction significantly, especially for the higher mass WDs.

Regarding the uncertainties in modelling the post-nova SSSs, there are two main elements relevant to our calculations. Firstly, it is the issue of mixing between the accreted matter with the WD core, the mechanism of which is still largely an unresolved issue in nova theory. Mixing is not accounted for in the MESA models of \citet{Wolf} that we have used, and therefore our results are insensitive to the WD core composition. Nevertheless, as evident from the analysis of the ejecta composition of observed novae, some level of mixing between the core and the envelope should occur in novae. The resulting metal enrichment in the nuclear burning shell enhances the burning rate of hydrogen during the CNO cycle, in turn increasing the burning temperature and also the kinetic energy of the ejecta \citep[e.g.,][]{Starrfield, Glasner-2012}. This produces a more violent nova eruption, reducing the remnant hydrogen mass that powers the SSS phase. The effect of mixing on the post-nova SSS phase is thus to shorten its duration. For example, \citet{Sala} have computed using hydrostatic nova envelope models, typical evolutionary timescales for the post-nova phase. They expressed this timescale in terms of the time interval required for $T_{\rm eff}$ to increase by 10~eV near the peak. We find the corresponding time interval from the MESA models to be larger by a factor between $\sim 10$ -- 20 than their models with the highest metal enrichment (ONe75 in their paper) and by a factor of $\lesssim 3$ than their models with the least metal enhancement (ONe25 in their paper). This indicates that mixing may be an important factor in explaining the discrepancy between our predictions and observation of the post-nova SSSs. Alternatively, observations of post-nova SSSs could be used to constrain the mixing in novae. For example, our results demonstrate that the {\it XMM-Newton} observation results would be incompatible with the most metal enhanced ONe75 model of \citet{Sala}, since the much shortened SSS phase from such metal enhancement would have reduced our predicted number of sources by about the same factor of $\sim10$ -- 20. On the other hand, their ONe25 model with less metal enhancement may be quite compatible with our calculations. Furthermore, for novae occurring on massive WDs with very short recurrence time, convection is weak and particle diffusion inefficient to allow for a strong mixing between the WD core and envelope, such that mixing may not be as significant as for the novae on less massive WDs.

Another open issue is the amount of circumstellar absorption produced by the nova ejecta itself or by the mass-loss from the system in general. In order to gauge the importance of this factor, we calculate the neutral hydrogen column density $\Delta N_{\rm H}$ required to reduce the predicted number of post-nova SSSs in Sec.~\ref{sec:obs_com} to the observed value of $\approx 16$. The results are listed in the last column of Table~\ref{tab:model_results}. As one can see, a moderate absorption by the  circumstellar material, at the level of $\sim (1-2)\times 10^{21}~{\rm cm}^{-2}$ is sufficient to reconcile our predictions with the observed number from \textit{XMM-Newton}. Since the detection of the post-nova SSSs is most likely around the time of their soft X-ray peak luminosities (Eq.~\eqref{eq:diff}, Fig.~\ref{fig:lcs}), the effect of circumstellar absorption is therefore most relevant around that time. There are, indeed, results from X-ray observations of novae that indicate the presence of circumstellar $N_{\rm H}$ around the time of their peak soft X-ray emission. For example, \citet{Page-2010} carried out spectral fitting of the X-ray observations of V2491~Cyg using blackbody model and \citet{Kahabka-1999}, that of U~Sco using NLTE model and obtained the evolution of $N_{\rm H}$ during their supersoft phase. They find the value of $N_{\rm H}$ for these systems around their peak emission time to be in excess from the interstellar medium by $\sim 10^{21}~{\rm cm}^{-2}$. Thus, we see that the $\Delta N_{\rm H}$ values that we have found (see Table~\ref{tab:model_results}) is quite compatible with the observed values.

Further, our predictions are  also dependent on the assumptions regarding the shape of the WD mass distribution in novae. One can see from Table~\ref{tab:model_results} that in terms of the total number of post-nova SSSs, the dependence is not very strong,  the largest difference being of a factor of  $\sim 2$. Therefore, our comparison of the predicted number of sources with observations is not strongly dependent on the assumed WD mass function in novae.  Dependence of the slope of the luminosity function and of the effective temperature distribution,  however, is more significant and can strongly affect the number (or probability of occurrence) of luminous and/or high temperature sources, by upto an order of magnitude in the extreme case. Thus, more detailed analysis of the luminosity function and temperature distribution of post-nova SSSs could, in principle, help in placing constraints on the mass distribution of WD in novae. This, of course, requires that other uncertainties of the models are dealt with.

\section{Summary and conclusions}\label{sec:conclude}

We have computed the expected population of post-nova SSSs in M31 using the multicycle nova evolutionary models of \citet{Wolf}, under various assumptions regarding the WD mass distribution in novae. For the latter, we consider the mass distributions obtained from the observed nova statistics in M31,  a simple theoretical distribution from \citet{Truran} and an ad hoc flat distribution. Our calculations predict that depending on the assumed WD mass function, there should be about ${\sim 250\mbox{--}600}$ post-nova SSSs in M31 with intrinsic (unabsorbed) luminosity in the $0.2-1.0$~keV band exceeding $10^{36}$~erg/s, and with combined luminosity of $\sim (2-4)\times 10^{39}$~erg/s (Table~\ref{tab:model_results}). Results obtained for different WD mass distributions are consistent with each other within a factor of $\sim 2$.  The  luminosity functions for all four WD mass distribution models exhibit a cut-off at $L_{\rm x}\approx 2\times 10^{38}$~erg/s, which corresponds to the maximum soft X-ray luminosity predicted by the post-nova evolutionary tracks. In a similar manner, the differential $T_{\rm eff}$ distributions for these models have a cut-off at $T_{\rm eff}\approx 1.5\times 10^{6}$~K, corresponding to the highest $T_{\rm eff}$ achieved by  the evolutionary tracks. 

We compare our predictions with the results of dedicated monitoring of the central part of M31 with \textit{XMM-Newton} \citep{Henze-1, Henze-2}. Using Monte-Carlo simulations and taking into account varying intrinsic absorption in M31 within the {\it XMM-Newton} FOV, we estimated the total number of post-nova SSSs which should have been detected by {\it XMM-Newton} in the course of these observations, to be of the order ${\approx 30-80}$. This exceeds by a factor of ${\approx 2-5}$ the actual  number of detected sources of $\approx 16$. Taking into account  the number and magnitude of uncertainties involved in modelling post-nova SSSs and possible incompleteness of the observed SSS statistics, we consider this as a good agreement. Furthermore, there are several uncertainties and simplifications in our calculations, which tend to overestimate the  number of SSSs. Most important among these are the following. The blackbody assumption used for the post-nova SSSs tends to overestimate the soft X-ray luminosity as compared to the NLTE WD atmosphere models. The mixing of the WD core material with the accreted envelope material can shorten the SSS phase and, correspondingly, reduce the number of post-nova SSSs. Finally, the presence of circumstellar material,  at an average level of $N_H\sim 10^{21}~{\rm cm}^{-2}$, would fully resolve the discrepancy.

\section*{Acknowledgments}
The authors would like to thank Pauline Barmby for providing the Spitzer 3.6 micron image of M31. We also thank Elly~M.~Berkhuijsen for providing the HI map of \citet{Brinks}. The TheoSSA service (http://dc.g-vo.org/theossa) used to retrieve theoretical spectra for this paper was constructed as part of the activities of the German Astrophysical Virtual Observatory. This work was supported by the National Science Foundation under grants PHY 11-25915, AST 11-09174, and AST 12-05574. Most of the MESA simulations for this work were made possible by the Triton Resource, a high-performance research computing system operated by the San Diego Supercomputer Center at UC San Diego. We are also grateful to Bill Paxton for his continual development of MESA. We also thank Martin~Henze for helpful comments on the {\em XMM-Newton} observations. We are grateful to John Grula, Dan Kohne and Francois Schweizer at the Carnegie Observatories for their help in searching for the exact observing dates of Arp's M31 survey. Finally, the authors would like to thank the referee Jan-Uwe Ness for the constructive comments and suggestions, which helped to improve the paper.

\bibliographystyle{mn2e}
\bibliography{ref_sss}

\begin{thebibliography}{56}
\expandafter\ifx\csname natexlab\endcsname\relax\def\natexlab#1{#1}\fi

\bibitem[{{Arp}(1956)}]{Arp}
{Arp} H.~C., 1956, \aj, 61, 15

\bibitem[{{Barmby} {et~al}\mbox{.}(2006){Barmby}, {Ashby}, {Bianchi},
  {Engelbracht}, {Gehrz}, {Gordon}, {Hinz}, {Huchra}, {Humphreys}, {Pahre},
  {P{\'e}rez-Gonz{\'a}lez}, {Polomski}, {Rieke}, {Thilker}, {Willner}, \&
  {Woodward}}]{Barmby}
{Barmby} P. {et~al.}, 2006, \apjl, 650, L45

\bibitem[{{Bell} \& {de Jong}(2001)}]{Bell}
{Bell} E.~F., {de Jong} R.~S., 2001, \apj, 550, 212

\bibitem[{{Bode} \& {Evans}(2008)}]{Bode}
{Bode} M.~F., {Evans} A., 2008, Classical Novae, 2nd Edition.~Edited by
  M.F.~Bode and A.~Evans~(Cambridge University Press), Cambridge Astrophysics
  Series, No.~43

\bibitem[{{Bogd{\'a}n} \& {Gilfanov}(2010)}]{Bogdan}
{Bogd{\'a}n} {\'A}., {Gilfanov} M., 2010, \mnras, 405, 209

\bibitem[{{Brinks} \& {Shane}(1984)}]{Brinks}
{Brinks} E., {Shane} W.~W., 1984, \aaps, 55, 179

\bibitem[{{Capaccioli} {et~al}\mbox{.}(1989){Capaccioli}, {della Valle},
  {Rosino}, \& {D'Onofrio}}]{Capaccioli}
{Capaccioli} M., {della Valle} M., {Rosino} L., {D'Onofrio} M., 1989, \aj, 97,
  1622

\bibitem[{{Darnley} {et~al}\mbox{.}(2004){Darnley}, {Bode}, {Kerins}, \& {et
  al.}}]{Darnley-2}
{Darnley} M.~J., {Bode} M.~F., {Kerins} E., {et al.}, 2004, \mnras, 353, 571

\bibitem[{{Darnley} {et~al}\mbox{.}(2006){Darnley}, {Bode}, {Kerins}, \& {et
  al.}}]{Darnley}
{Darnley} M.~J., {Bode} M.~F., {Kerins} E., {et al.}, 2006, \mnras, 369, 257

\bibitem[{{Denissenkov} {et~al}\mbox{.}(2013){Denissenkov}, {Herwig},
  {Bildsten}, \& {Paxton}}]{Denissenkov-2013}
{Denissenkov} P.~A., {Herwig} F., {Bildsten} L., {Paxton} B., 2013, \apj, 762,
  8

\bibitem[{{Fujimoto}(1982{\natexlab{a}})}]{Fujimoto-1982b}
{Fujimoto} M.~Y., 1982{\natexlab{a}}, \apj, 257, 767

\bibitem[{{Fujimoto}(1982{\natexlab{b}})}]{Fujimoto-1982a}
{Fujimoto} M.~Y., 1982{\natexlab{b}}, \apj, 257, 752

\bibitem[{{Glasner}, {Livne} \& {Truran}(2012){Glasner}, {Livne}, \&
  {Truran}}]{Glasner-2012}
{Glasner} S.~A., {Livne} E., {Truran} J.~W., 2012, \mnras, 427, 2411

\bibitem[{{Henze} {et~al}\mbox{.}(2014{\natexlab{a}}){Henze}, {Ness},
  {Darnley}, {Bode}, {Williams}, {Shafter}, {Kato}, \& {Hachisu}}]{Henze-2014}
{Henze} M., {Ness} J.-U., {Darnley} M.~J., {Bode} M.~F., {Williams} S.~C.,
  {Shafter} A.~W., {Kato} M., {Hachisu} I., 2014{\natexlab{a}}, \aap, 563, L8

\bibitem[{{Henze} {et~al}\mbox{.}(2015){Henze}, {Ness}, {Darnley}, {Bode},
  {Williams}, {Shafter}, {Sala}, {Kato}, {Hachisu}, \& {Hernanz}}]{Henze-2015}
{Henze} M. {et~al.}, 2015, ArXiv e-prints

\bibitem[{{Henze} {et~al}\mbox{.}(2013){Henze}, {Pietsch}, {Haberl}, {Della
  Valle}, {Riffeser}, {Sala}, {Hatzidimitriou}, {Hofmann}, {Hartmann},
  {Koppenhoefer}, {Seitz}, {Williams}, {Hornoch}, {Itagaki}, {Kabashima},
  {Nishiyama}, {Xing}, {Lee}, {Magnier}, \& {Chambers}}]{Henze-2013}
{Henze} M. {et~al.}, 2013, \aap, 549, A120

\bibitem[{{Henze} {et~al}\mbox{.}(2014{\natexlab{b}}){Henze}, {Pietsch},
  {Haberl}, {Della Valle}, {Sala}, {Hatzidimitriou}, {Hofmann}, {Hernanz},
  {Hartmann}, \& {Greiner}}]{Henze-3}
{Henze} M. {et~al.}, 2014{\natexlab{b}}, \aap, 563, A2

\bibitem[{{Henze} {et~al}\mbox{.}(2010){Henze}, {Pietsch}, {Haberl}, {Hernanz},
  {Sala}, {Della Valle}, {Hatzidimitriou}, {Rau}, {Hartmann}, {Greiner},
  {Burwitz}, \& {Fliri}}]{Henze-1}
{Henze} M. {et~al.}, 2010, \aap, 523, A89

\bibitem[{{Henze} {et~al}\mbox{.}(2011){Henze}, {Pietsch}, {Haberl}, {Hernanz},
  {Sala}, {Hatzidimitriou}, {Della Valle}, {Rau}, {Hartmann}, \&
  {Burwitz}}]{Henze-2}
{Henze} M. {et~al.}, 2011, \aap, 533, A52

\bibitem[{{Jos{\'e}} \& {Hernanz}(1998)}]{Jose}
{Jos{\'e}} J., {Hernanz} M., 1998, \apj, 494, 680

\bibitem[{{Kahabka} {et~al}\mbox{.}(1999){Kahabka}, {Hartmann}, {Parmar}, \&
  {Negueruela}}]{Kahabka-1999}
{Kahabka} P., {Hartmann} H.~W., {Parmar} A.~N., {Negueruela} I., 1999, \aap,
  347, L43

\bibitem[{{Kato} \& {Hachisu}(1994)}]{Kato-1994}
{Kato} M., {Hachisu} I., 1994, \apj, 437, 802

\bibitem[{{Krautter}(2002)}]{Krautter}
{Krautter} J., 2002, in American Institute of Physics Conference Series, Vol.
  637, Classical Nova Explosions, {Hernanz} M., {Jos{\'e}} J., eds., pp.
  345--354

\bibitem[{{Ness}(2010)}]{Ness-2010}
{Ness} J.-U., 2010, Astronomische Nachrichten, 331, 179

\bibitem[{{Ness} {et~al}\mbox{.}(2013){Ness}, {Osborne}, {Henze}, {Dobrotka},
  {Drake}, {Ribeiro}, {Starrfield}, {Kuulkers}, {Behar}, {Hernanz}, {Schwarz},
  {Page}, {Beardmore}, \& {Bode}}]{Ness}
{Ness} J.-U. {et~al.}, 2013, \aap, 559, A50

\bibitem[{{Nieten} {et~al}\mbox{.}(2006){Nieten}, {Neininger}, {Gu{\'e}lin},
  {Ungerechts}, {Lucas}, {Berkhuijsen}, {Beck}, \& {Wielebinski}}]{Nieten}
{Nieten} C., {Neininger} N., {Gu{\'e}lin} M., {Ungerechts} H., {Lucas} R.,
  {Berkhuijsen} E.~M., {Beck} R., {Wielebinski} R., 2006, \aap, 453, 459

\bibitem[{{Page} {et~al}\mbox{.}(2010){Page}, {Osborne}, {Evans}, {Wynn},
  {Beardmore}, {Starling}, {Bode}, {Ibarra}, {Kuulkers}, {Ness}, \&
  {Schwarz}}]{Page-2010}
{Page} K.~L. {et~al.}, 2010, \mnras, 401, 121

\bibitem[{{Paxton} {et~al}\mbox{.}(2011){Paxton}, {Bildsten}, {Dotter},
  {Herwig}, {Lesaffre}, \& {Timmes}}]{Paxton-2011}
{Paxton} B., {Bildsten} L., {Dotter} A., {Herwig} F., {Lesaffre} P., {Timmes}
  F., 2011, \apjs, 192, 3

\bibitem[{{Paxton} {et~al}\mbox{.}(2013){Paxton}, {Cantiello}, {Arras},
  {Bildsten}, {Brown}, {Dotter}, {Mankovich}, {Montgomery}, {Stello}, {Timmes},
  \& {Townsend}}]{Paxton-2013}
{Paxton} B. {et~al.}, 2013, \apjs, 208, 4

\bibitem[{{Pietsch} {et~al}\mbox{.}(2005){Pietsch}, {Fliri}, {Freyberg},
  {Greiner}, {Haberl}, {Riffeser}, \& {Sala}}]{Pietsch-2005}
{Pietsch} W., {Fliri} J., {Freyberg} M.~J., {Greiner} J., {Haberl} F.,
  {Riffeser} A., {Sala} G., 2005, \aap, 442, 879

\bibitem[{{Pietsch} {et~al}\mbox{.}(2007){Pietsch}, {Haberl}, {Sala}, {Stiele},
  {Hornoch}, {Riffeser}, {Fliri}, {Bender}, {B{\"u}hler}, {Burwitz}, {Greiner},
  \& {Seitz}}]{Pietsch-2007}
{Pietsch} W. {et~al.}, 2007, \aap, 465, 375

\bibitem[{{Prialnik}(1986)}]{Prialnik-1986}
{Prialnik} D., 1986, \apj, 310, 222

\bibitem[{{Prialnik} \& {Kovetz}(1995)}]{Prialnik-1995}
{Prialnik} D., {Kovetz} A., 1995, \apj, 445, 789

\bibitem[{{Rauch} {et~al}\mbox{.}(2010){Rauch}, {Orio}, {Gonzales-Riestra},
  {Nelson}, {Still}, {Werner}, \& {Wilms}}]{Rauch-2010b}
{Rauch} T., {Orio} M., {Gonzales-Riestra} R., {Nelson} T., {Still} M., {Werner}
  K., {Wilms} J., 2010, \apj, 717, 363

\bibitem[{{Rauch} \& {Werner}(2010)}]{Rauch-2010}
{Rauch} T., {Werner} K., 2010, Astronomische Nachrichten, 331, 146

\bibitem[{{Sala} \& {Hernanz}(2005{\natexlab{a}})}]{Sala-b}
{Sala} G., {Hernanz} M., 2005{\natexlab{a}}, \aap, 439, 1057

\bibitem[{{Sala} \& {Hernanz}(2005{\natexlab{b}})}]{Sala}
{Sala} G., {Hernanz} M., 2005{\natexlab{b}}, \aap, 439, 1061

\bibitem[{{Salpeter}(1955)}]{Salpeter}
{Salpeter} E.~E., 1955, \apj, 121, 161

\bibitem[{{Skopal}(2015)}]{Skopal}
{Skopal} A., 2015, \na, 36, 116

\bibitem[{{Soraisam} \& {Gilfanov}(2014)}]{Soraisam}
{Soraisam} M., {Gilfanov} M., 2014, ArXiv e-prints

\bibitem[{{Sparks}, {Starrfield} \& {Truran}(1976){Sparks}, {Starrfield}, \&
  {Truran}}]{Starrfield-3}
{Sparks} W.~M., {Starrfield} S., {Truran} J.~W., 1976, \apj, 208, 819

\bibitem[{{Stark} {et~al}\mbox{.}(1992){Stark}, {Gammie}, {Wilson}, {Bally},
  {Linke}, {Heiles}, \& {Hurwitz}}]{Stark}
{Stark} A.~A., {Gammie} C.~F., {Wilson} R.~W., {Bally} J., {Linke} R.~A.,
  {Heiles} C., {Hurwitz} M., 1992, \apjs, 79, 77

\bibitem[{{Starrfield}, {Sparks} \& {Truran}(1974{\natexlab{a}}){Starrfield},
  {Sparks}, \& {Truran}}]{Starrfield-2}
{Starrfield} S., {Sparks} W.~M., {Truran} J.~W., 1974{\natexlab{a}}, \apjs, 28,
  247

\bibitem[{{Starrfield}, {Sparks} \& {Truran}(1974{\natexlab{b}}){Starrfield},
  {Sparks}, \& {Truran}}]{Starrfield-2b}
{Starrfield} S., {Sparks} W.~M., {Truran} J.~W., 1974{\natexlab{b}}, \apj, 192,
  647

\bibitem[{{Starrfield} {et~al}\mbox{.}(1972){Starrfield}, {Truran}, {Sparks},
  \& {Kutter}}]{Starrfield}
{Starrfield} S., {Truran} J.~W., {Sparks} W.~M., {Kutter} G.~S., 1972, \apj,
  176, 169

\bibitem[{{Tang} {et~al}\mbox{.}(2014){Tang}, {Bildsten}, {Wolf}, {Li}, {Kong},
  {Cao}, {Cenko}, {De Cia}, {Kasliwal}, {Kulkarni}, {Laher}, {Masci}, {Nugent},
  {Perley}, {Prince}, \& {Surace}}]{Tang-2014}
{Tang} S. {et~al.}, 2014, \apj, 786, 61

\bibitem[{{Townsley} \& {Bildsten}(2004)}]{Townsley-2004}
{Townsley} D.~M., {Bildsten} L., 2004, \apj, 600, 390

\bibitem[{{Townsley} \& {Bildsten}(2005)}]{Townsley-2005}
{Townsley} D.~M., {Bildsten} L., 2005, \apj, 628, 395

\bibitem[{{Truran} \& {Livio}(1986)}]{Truran}
{Truran} J.~W., {Livio} M., 1986, \apj, 308, 721

\bibitem[{{Tuchman} \& {Truran}(1998)}]{Tuchman}
{Tuchman} Y., {Truran} J.~W., 1998, \apj, 503, 381

\bibitem[{{van Rossum}(2012)}]{Rossum-2012}
{van Rossum} D.~R., 2012, \apj, 756, 43

\bibitem[{{van Rossum} \& {Ness}(2010)}]{Rossum-2010}
{van Rossum} D.~R., {Ness} J.-U., 2010, Astronomische Nachrichten, 331, 175

\bibitem[{{Walterbos} \& {Kennicutt}(1987)}]{Walterbros}
{Walterbos} R.~A.~M., {Kennicutt}, Jr. R.~C., 1987, \aaps, 69, 311

\bibitem[{{Wilms}, {Allen} \& {McCray}(2000){Wilms}, {Allen}, \&
  {McCray}}]{Wilms}
{Wilms} J., {Allen} A., {McCray} R., 2000, \apj, 542, 914

\bibitem[{{Wolf} {et~al}\mbox{.}(2013){Wolf}, {Bildsten}, {Brooks}, \&
  {Paxton}}]{Wolf}
{Wolf} W.~M., {Bildsten} L., {Brooks} J., {Paxton} B., 2013, \apj, 777, 136

\bibitem[{{Yaron} {et~al}\mbox{.}(2005){Yaron}, {Prialnik}, {Shara}, \&
  {Kovetz}}]{Yaron}
{Yaron} O., {Prialnik} D., {Shara} M.~M., {Kovetz} A., 2005, \apj, 623, 398

\end{thebibliography}

\appendix
\section{Incompleteness correction of Arp's survey of M31 (1953-1955)}
\label{append:Arp}

We apply the method developed by \citet{Soraisam} for an approximate incompleteness calculation of Arp's survey of M31 (1953-1955) \citep{Arp}. We carry out Monte-Carlo simulations taking into account the limiting magnitude and the observing pattern of the survey, and by using a scalable template nova light curve and the observed maximum magnitude -- rate of decline relation from \citet{Soraisam}. We then compute the fraction of detected novae as a function of the decline time $t_{2}$.

Arp's survey of M31 spanned two seasons between June 1953 and January 1955, and 30 novae were discovered during its course \citep{Arp}. The survey was carried out using the 60-inch telescope on Mount Wilson, equipped with photographic plates of size 5x7 inch at the focal plane with a plate scale of 24~arcsec~per~mm. The field of view (FOV) thus covered $\sim1$~sq.~degree of the galaxy and the average limiting magnitude was $m_{pg}=20$ \citep{Capaccioli}. The FOV was centered primarily at two points located $10'$ on either side of the galactic nucleus along the major axis. Some addtional exposures were also taken, which were centered on the galactic nucleus as well as two points located $20'$ either side of the nucleus along the major axis. In our calculations, we have considered the primary exposures (which also include all the 30 detected novae in this survey) and the exposures centered on the nucleus, excluding a circular region of radius $2'$ in the innermost part, where this survey is known to be incomplete \citep{Capaccioli}.

\begin{figure}
\includegraphics[width=\columnwidth]{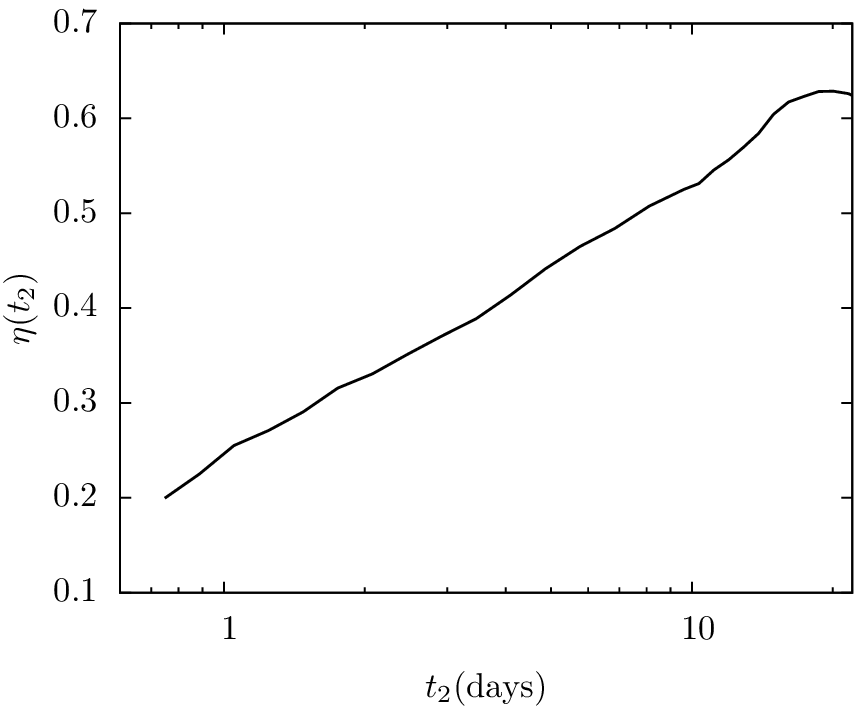}
\caption{The detection efficiency $\eta$ of Arp's survey of M31 (1953-1955), approximately computed using the observation details provided in the original Arp's publication and the online database of the Carnegie Observatories' plate archive, as described in Appendix~\ref{append:Arp}. The detection efficiency is the fraction of novae with the given  light curve decay time $t_2$ detected in the survey.}\label{fig:arp}
\end{figure}

The observing patterns for this survey are not described in the original Arp's publication \citet{Arp}. However, the photographic plates from Arp's survey are in the archive of the Carnegie Observatories and the details of the plates have been cataloged in their online database{\footnote{\label{PAST}\url{http://plates.obs.carnegiescience.edu/PAST/search/}}}. From this database, we obtain the exact observing dates. The observing times for multiple plates taken on the same night are, however, not available. Nevertheless, \citet{Arp} has described the observations each night to be repeated every hour or so, and therefore we assume the same for the plate entries in the database. The resulting completeness curve of Arp's survey is shown in Fig.~\ref{fig:arp} as a function of the light curve decay time $t_2$.  We should note that the above calculations have not accounted for the spatial variations in the internal optical extinction in M31. However, extinction variations should not have a significant effect on their detection efficiency, as shown by \citet{Soraisam}. 
Furthermore, the detection of novae by Arp was not automated and it involved some human intervention, as was the case for all pre-CCD surveys. Therefore, an accurate calculation of the completeness of Arp's catalog is impossible. However, the approximate completeness computed here is a reasonable first approximation and allows us to  estimate the population of fast novae in M31 (see below). The existing high cadence surveys such as the iPTF and Pan-STARRS, would allow a more efficient detection of fast novae. A more precise determination of their population in M31 then should be and will be performed using these modern facilities.

\section{Nova rate distribution in M31}
\label{append:rate_dist}

In order to obtain the $t_2$ distribution of novae in M31 we will combine the data of the Arp's survey with the nova catalogue of \citet{Darnley-2}.

\begin{figure}
\includegraphics[width=\columnwidth]{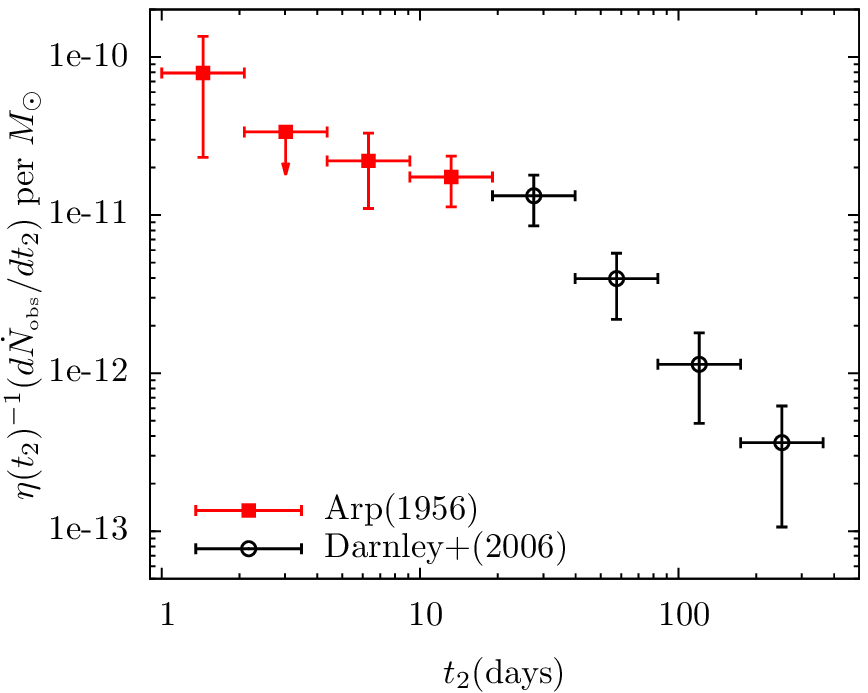}
\caption{ Observed differential nova rate distribution in M31 obtained using the nova sample of \citet{Arp} for $t_{2}\leq 20$~days and that of \citet{Darnley-2} for $t_{2}>20$~days, after correcting for incompleteness (see Appendix~\ref{append:rate_dist} for details).}\label{fig:arp_darnley_rate}
\end{figure}

For the nova sample of \citet{Darnley-2} obtained from the POINT-AGAPE (Pixel-lensing Observations with the Isaac Newton Telescope -- Andromeda Galaxy Amplified Pixels Experiment) survey of M31, \citet{Darnley} carried out a comprehensive analysis of the spatial completeness of the survey. However, this catalogue is not complete for fast novae,  due to insufficient cadence of the POINT-AGAPE survey \citep{Darnley}. We therefore combine the  Arp's sample of fast novae  ($t_{2}\leq 20$~days), corrected for incompleteness using the results in Appendix~\ref{append:Arp}, and the nova sample of \citet{Darnley-2} for $t_{2}>20$~days, using the completeness value of $\approx 25\%$ obtained from Table 3 of \citet{Darnley} for correction. In particular, we use their $\theta=1$ case.  It is still an open question whether the specific nova rates (per unit mass) in the bulge and disk  of M31 differ much.  \citet{Darnley} have shown that the overall nova rate in M31 derived from their data does not depend significantly on their  ratio. Therefore,  we  assumed for simplicity  that specific  nova rates in the bulge and disk are same  (i.e., $\theta=1$ case of \citealt{Darnley}).

The nova samples of \citet{Arp} and \citet{Darnley-2} are then normalized by the stellar mass contained in the FOV of the corresponding survey. For Arp's survey, we estimate the mass enclosed by the FOV using the Spitzer 3.6~micron mosaic image of M31 from \citet{Barmby}. For this image, which is also background subtracted, there is insignificant contamination from foreground and background objects; these contribute only $\lesssim5\%$ to the luminosity within the FOV. Assuming $K-[3.6]=0.3$ as done in \citet{Barmby} (see also \citealt{Bogdan}), and using the K-band mass-to-light ratio of 0.80 solar units derived from \citet{Bell} using the color of M31, $B-R=1.5$ \citep{Walterbros}, we thus compute the stellar mass within the FOV of the survey, which is $\approx6\times10^{10}M_\odot$. For the POINT-AGAPE survey, the mass is estimated using the fraction of stellar light enclosed by the survey FOV from Table 3 in \citet{Darnley}, which gives us a value of $\approx4\times10^{10}~M_\odot$. The resulting combined  $t_2$ distribution of the incompleteness corrected observed nova rate in M31, normalized to unit $M_\odot$, is shown in Fig.~\ref{fig:arp_darnley_rate}. Its total rate is ${\approx 106~{\rm yr}^{-1}}$; this value is larger than $75~{\rm yr}^{-1}$ derived by \citet{Darnley} for their $\theta=1$ model because of the contribution of the fast novae.

\label{lastpage}

\end{document}